\documentclass[12pt,draftcls,onecolumn]{IEEEtran}

\ifCLASSINFOpdf
\else
\fi
\usepackage{graphics} 
\usepackage{epsfig} 
\usepackage{epstopdf}
\usepackage{amsmath} 
\usepackage{amssymb}  

\usepackage{cite}
\usepackage{graphicx}
\usepackage{pgf}
\usepackage{bbm}
\usepackage{etoolbox}
\usepackage{mathrsfs}
\usepackage{enumerate}
\usepackage{color}
\usepackage[textsize=footnotesize,textwidth=1.5cm]{todonotes}	

\newtheorem{thm}{Theorem}
\newtheorem{assum}{Assumption}
\newtheorem{lem}[thm]{Lemma}
\newtheorem{definition}{Definition}
\newtheorem{rem}{Remark}

\usepackage[normalem]{ulem}
\newcommand{\argmin}{\operatornamewithlimits{argmin}}

\newtoggle{jou}
\togglefalse{jou}

\begin{document}
%
\title{Dynamic Collective Choice:
Social Optima}
%
%
%

\author{Rabih~Salhab, Jerome~Le~Ny              
        and~Roland~P.~Malham\'e
\thanks{This work was supported by NSERC under Grants 6820-2011 and 435905-13.
The authors are with the department of Electrical Engineering, 
Polytechnique Montreal and with GERAD, Montreal, QC H3T-1J4, Canada {\tt\small \{rabih.salhab, jerome.le-ny, roland.malhame\}@polymtl.ca}}}

%
%

\markboth{}%
{Rabih \MakeLowercase{\textit{et al.}}: Bare Demo of IEEEtran.cls for Journals}
%



\maketitle

\begin{abstract}
We consider a dynamic 
collective choice problem where a large number of 
players are
cooperatively choosing between multiple destinations 
while being influenced by the behavior of the group. 
For example, in a robotic swarm exploring a new environment,
a robot might have to choose between multiple sites to visit, but
at the same time it should remain close to the group to achieve some coordinated tasks. 
We show that to find a social optimum for our problem, one needs
to solve a set of Linear Quadratic Regulator
problems, whose number  
increases exponentially with the size of the population. Alternatively, we develop 
via the Mean Field Games methodology a set of decentralized strategies 
that are independent of the size of the population. When the number of agents 
is sufficiently large, these strategies qualify as approximately socially optimal. 
To compute the approximate social optimum, each player needs to know its own 
state and the statistical distributions of the players' initial states and 
problem parameters. Finally, we give a  
numerical example where the cooperative 
and non-cooperative cases have 
opposite behaviors. Whereas in the former
the size of the majority increases with the 
social effect, in the latter, the existence 
of a majority is disadvantaged.
\end{abstract}


%
\IEEEpeerreviewmaketitle

\section{Introduction}

Discrete choice models were developed in economics to understand 
human choice behavior. A concern of these models is predicting 
the decision of an individual in face of a set of alternatives, 
for example, anticipating a traveler's choice between different modes 
of transportation \cite{Kappelman2005}. 
These choices depend on some personal characteristics, such as 
the traveler's financial situation, on some attributes of the 
alternatives, such as their prices, and on some unobservable 
attributes, e.g., the traveler's taste. 
The first static discrete choice model was proposed by 
McFadden in \cite{mcfadden1973conditional}. 

In some situations, the individuals' choices are socially
influenced, that is, an individual's choice is affected by the 
others' choices, for example entry or withdrawal from the labor 
market in cooperative families \cite{bresnahan1991empirical}. 
The main goal of this paper is to model within the framework 
of dynamic \emph{cooperative} game theory situations
where a large number of players/agents are making socially influenced 
choices among a finite set of alternatives. The players involved
in this game are weakly coupled, that is, the individual choices 
are considerably influenced by a functional of the others' choice
distribution (in this paper the mean), but for a sufficiently large 
population, an isolated individual's choice has a negligible influence 
on the others' choices. Moreover, the players' states contributing to 
the social effect are assumed indistinguishable. 
In navigation applications for example, a planner might want to
deploy a swarm of robots cooperating to explore an unknown terrain. 
A robot faces a situation where it should choose between multiple 
sites to visit. At the same time, it should remain closed to the
group to achieve some coordinated tasks 
\cite{Nourian11_collective, DistCtrlRobotNetw, jerome13_ada}.

In non-cooperative games, perfectly rational players act 
selfishly by minimizing their individual costs irrespective of
making the other players better off or worse off. 
This ``utilitarianist" aspect of non-cooperative games neglects 
the social context where the social norms, social values, 
the presence of a social planner or the social structures 
impose a kind of cooperation between the players. 
An example of the influence of the social context on the
behavior of players was given in \cite{EconBiz-10001307135}, 
where the author shows how at the Chicago Options Exchange 
the relations among the traders, supposed to be noncooperative, 
affect their trades. 
In the robotic swarm example, the cooperative behavior of the robots results from the intention of the planner to optimize a total cost.
Whereas in the non-cooperative case the agents search for a Nash 
equilibrium, the players seek in the cooperative case
a totally
different type of solution, namely a social optimum. 

The Mean Field Games (MFG) methodology, 
which we follow in this paper, 
is concerned with dynamic games
involving a large number of weakly 
coupled agents.
It was originally developed
in a series of papers to study dynamic non-cooperative 
games \cite{Huang03_wirelessPower, Huang06_particles, Huang07_Large, 
huang2006large, lasry2006jeux, lasry2006jeux2, Lasry07_MFG}. 
The cooperative Linear Quadratic Gaussian (LQG) MFG formulation 
was developed later in \cite{huang2012social}, where the 
authors investigate the structure of the LQG costs to develop 
for a continuum of agents a set of decentralized person-by-person 
optimal strategies (a weaker solution concept than the social optimum 
that coincides under some conditions with the social optimum 
\cite{ho1972team, yuksel2013stochastic}). Moreover, they show that 
these strategies, when applied by a finite population, converge to an 
exact social optimum as the number of players increases to infinity.

The main contribution of this paper is as follows. We consider a 
cooperative collective choice model where the number of candidate 
optimal control laws  increases exponentially with the size of 
the population. Then, we develop a set of decentralized   
strategies of dimensions independent of the size of the population and that
converge to the social optimum as the size of the population 
increases to infinity. Although the methodology used to solve the 
game follows \cite{huang2012social}, the non-smoothness and 
non-convexity of our final costs, which involve a minimum function,
require different proofs for the convergence of the mean field based 
decentralized strategies to the social optimum, see Lemmas \ref{appr}, 
\ref{2-app}, Theorem \ref{th-optim} and Remark \ref{need_L_p}.
%
%
In particular, our problem formulation results in decentralized strategies 
that are \textit{discontinuous} with respect to the agents' initial conditions,
capturing the issue of choosing between a finite set of alternatives, which
cannot be modeled using the standard LQG MFG setup considered in \cite{huang2012social}.

In \cite{DBLP:journals/corr/SalhabMN15}, we studied the non-cooperative 
version of our model and developed via the MFG methodology approximate Nash strategy
profiles that converge to exact Nash equilibria as the number of players increases 
to infinity. Since the person-by-person solutions are Nash-like  solutions,
we rely in this note on some results established in \cite{DBLP:journals/corr/SalhabMN15} 
to establish the existence of the person-by-person solutions and compute them.
A static discrete choice model with social interactions was 
also studied by Brock and Durlauf in \cite{Brock2001}, where the 
authors develop a non-cooperative and a cooperative game 
involving a large number of players. Each player makes 
a choice between two alternatives while being affected by the 
average of its peers' decisions. Inspired by the statistical 
mechanics approach, Brock and Durlauf propose a methodology 
to solve the game that is similar to the MFG approach.


The cooperative dynamic discrete choice model is formulated 
in Section \ref{model}. We show in Section \ref{Centralized 
Pareto} that to find an exact social optimum, one can naively 
solve $l^N$ Linear Quadratic Regulator (LQR) problems, each of dimensions $Nn$, where $l$ 
is the number of choices, $N$ the number of players, 
and $n$ the dimension of the individual state spaces. 
Alternatively, we develop in Section \ref{decen} via the 
MFG approach and within the so-called person to person optimization setting 
a set of decentralized strategies that are asymptotically socially optimal. 
The dimensions of the decentralized strategies are independent of the size of 
the population. In Section \ref{sim} we give some simulation results, while
Section \ref{conclusion} presents our conclusions.


\section{Mathematical Model} \label{model}

We consider a cooperative game model 
involving $N$ players with 
linear dynamics
\begin{align} \label{eq-dynamics}
\dot{x}_i = A_ix_i+B_iu_i && 
i=1,\dots,N,
\end{align}
where $A_i \in \mathbb{R}^{n \times n}$, 
$B_i \in \mathbb{R}^{n \times m}$, 
$x_i \in \mathbb{R}^n$ is the state of 
agent $i$, $u_i \in 
U=L_2([0,T],\mathbb{R}^m)$ 
its control input and $x_i^0$ its 
initial state.  
The players cooperate to minimize a common social cost 
\begin{equation} \label{eq-social cost}
J_{soc}\left(u_1,\dots,u_N,x^{(N)}\right)=\sum_{i=1}^N 
J_i\left(u_i,x^{(N)}\right),
\end{equation}
where
\iftoggle{jou}{
\begin{multline} \label{eq-cost}
J_i\left(u_i,x^{(N)}\right)=\int_0^T 
 \left \{ \frac{q}{2} \left \|x_i-Z 
 x^{(N)}\right\|^2 + 
 \frac{r_i}{2} \|u_i\|^2 \right \} 
 \mathrm{dt} \\
 +\min\limits_{j=1,\dots,l}\left\{\frac{M_{ij}}
 {2} \|x_i(T)-p_j\|^2\right\} 
\end{multline}}{
\begin{equation} \label{eq-cost}
J_i\left(u_i,x^{(N)}\right)=\int_0^T 
 \left \{ \frac{q}{2} \left \|x_i-Z 
 x^{(N)}\right\|^2 + 
 \frac{r_i}{2} \|u_i\|^2 \right \} 
 \mathrm{dt} 
 +\min\limits_{j=1,\dots,l}\left\{\frac{M_{ij}}
 {2} \|x_i(T)-p_j\|^2\right\}
 \end{equation}
} 
are the individual costs, $q,r_i,M_{ij}>0$, $Z \in \mathbb{R}^{n \times n}$, 
and $p_j \in \mathbb{R}^n$, $j=1,\dots,l$, are the destination points. 
The individual cost functions penalize along the path the effort 
and the deviation from the mean. Moreover, each agent must be close 
at time $T$ to one of the destination points. Otherwise, it is strongly penalized by the final cost. 
The agents are cost coupled via the average $x^{(N)}= \frac{1}{N} \sum_{i=1}^N x_i$. 
The coefficient $r_i$ depends on
the agent $i$. 
In the robotic swarm example, this reflects, for instance, the intention of the 
social planner to limit the mobility of some robots. We assume that the
coefficient $M_{ij}$ depends on the agent $i$
and the destination point $p_j$ to impose initial preferences towards the alternatives, 
as discussed later in Remark \ref{initial tendency}.
When considering the limiting population ($N \to \infty$), it is convenient
to represent the limiting sequences of 
$(\theta_i)_{i=1,\dots,N}:=\{(A_i,B_i,r_i,M_{i1}, \dots, M_{il})\}_{i=1,\dots,N}$ 
and $\{x_i^0\}_{i=1,\dots,N}$ by two independent random variables $\theta$ and $x^0$ 
on some probability space $(\Omega, \mathcal{F}, \mathbb{P})$. 
We assume that $\theta$ is in a compact set $\Theta$.  
Let us denote the empirical measures of the sequences $\theta_i$ and $x_i^0$,    
$\mathbb{P}^N_\theta(\mathcal A) = 
\frac{1}{N} \sum_{i=1}^N 1_{\{\theta_i \in \mathcal A \}}$ 
and $\mathbb{P}^N_0(\mathcal A) = \frac{1}{N} \sum_{i=1}^N 1_{\{x^0_i \in \mathcal A\}}$  
for all (Borel) measurable sets $\mathcal A$. 
We assume that $\mathbb{P}^N_\theta$ and $\mathbb{P}^N_0$ have weak limits $\mathbb{P}_\theta$ 
and $\mathbb{P}_0$. For further discussions about this assumption, one can refer 
to \cite{huang2012social}.

A social optimum is defined as the optimal control law $(u_1^*,\dots,u_N^*)$ 
of (\ref{eq-social cost}). We start in the following section by solving for such
a social optimum in a centralized manner.

\section{Centralized Social Optimum} \label{Centralized Pareto}

In this section, we assume that each player can observe the 
states and the parameters of the other players. We define $x=(x_1,\dots,x_N)^T$ the 
state of the population and $u=(u_1,\dots,u_N)^T$ its strategy profile. 
The population's dynamics is then
\begin{equation} \label{eq-dynamics-population}
\dot{x}=\tilde{A}x+\tilde{B}u,
\end{equation}
where $\tilde{A} = \text{diag} (A_1,\dots,A_N)$ and $\tilde{B} = \text{diag} (B_1,\dots,B_N)$.
The individual costs can be written
\begin{equation} \label{ind_cost_min}
J_i\left(u_i,x^{(N)}\right)=\min\limits_{p_j \in \Delta} J_i^{p_j}\left(u_i,x^{(N)}\right),
\end{equation}
where $\Delta = \{ p_1,\dots,p_l \}$ and
\iftoggle{jou}{
\begin{multline*}
J_i^{p_j}\left(u_i,x^{(N)}\right)=\int_0^T 
 \left \{ \frac{q}{2} \left \|x_i-Z x^{(N)} \right\|^2 + 
 \frac{r_i}{2} \|u_i\|^2 \right \} \mathrm{dt} \\
 +\frac{M_{ij}}{2}\|x_i(T)-p_j\|^2.
\end{multline*}}{
\begin{equation*}
J_i^{p_j}\left(u_i,x^{(N)}\right)=\int_0^T 
 \left \{ \frac{q}{2} \left \|x_i-Z x^{(N)} \right\|^2 + 
 \frac{r_i}{2} \|u_i\|^2 \right \} \mathrm{dt} 
 +\frac{M_{ij}}{2}\|x_i(T)-p_j\|^2.
\end{equation*}
}
Using the equality $a+\min(b,c)=\min(a+b,a+c)$, one can prove by induction that 
the social cost (\ref{eq-social cost}) can be written
\begin{equation*}
J_{soc}\left(u,x^{(N)}\right) =
\min\limits_{d=(d_1,\dots,d_N) \in 
\Delta^N}
 \sum_{i=1}^NJ_i^{d_i}\left(u_i,x^{(N)}\right).
\end{equation*}
Noting that 
\begin{equation*}
\inf\limits_{u\in U^N} J_{soc}
\left(u,x^{(N)}\right)  = 
\min\limits_{d \in 
\Delta^N} 
\inf\limits_{u\in U^N} \sum_{i=1}^NJ_i^{d_i}\left(u_i,x^{(N)}
\right),
\end{equation*}
one can optimize the $l^N$ costs 
$J^d(u)=\sum_{i=1}^NJ_i^{d_i}(u_i,x^{(N)})$ and choose the less
costly
combination of destination points
$d^* \in \Delta^N$ which corresponds to the minimum of the optima of $J^d$.
The costs $J^d$, for $d \in \Delta^N$, can be written 
\iftoggle{jou}{
\begin{multline} \label{partial-cost}
J^d(u) = \int_0^T \left \{\frac{1}{2} x^T\tilde{Q}x + 
\frac{1}{2} u^T \tilde{R} u \right\} \mathrm{dt} + 
\\ \frac{1}{2}(x(T)-d)^T \tilde{M}^{d}(x(T)-d),
\end{multline}}{
\begin{equation} \label{partial-cost}
J^d(u) = \int_0^T \left \{\frac{1}{2} x^T\tilde{Q}x + 
\frac{1}{2} u^T \tilde{R} u \right\} \mathrm{dt} + 
\frac{1}{2}(x(T)-d)^T \tilde{M}^{d}(x(T)-d),
\end{equation}
}
where $\tilde{Q}=I_n\otimes I_N + \frac{1}{N}(11^T)\otimes L$, $\tilde{R} = \text{diag} (r_1I_m,\dots,r_NI_m)$, 
$\tilde{M}^d = \text{diag} (M_{1d_1}I_n,\dots,M_{Nd_N}I_n)$, and
\begin{equation} \label{L-eq}
L=Z^T Z -Z - Z^T,
\end{equation}
with $\otimes$ denoting the Kronecker product, $1=[1,\dots,1]^T$,
$\text{diag}(.)$ denoting a block diagonal matrix.

The LQR problem defined by (\ref{partial-cost}) and (\ref{eq-dynamics-population}) 
has a unique optimal control law \cite{anderson2007optimal}
\begin{equation} \label{cen-ctr}
u^{d}_*(t)= - \tilde{R}^{-1}
 \tilde{B}^T \left(\tilde{\Gamma}^d(t) x 
+ \tilde{\beta}^d(t)\right )
\end{equation}
with the corresponding optimal cost
\begin{equation} \label{cen-cost}
J^{d}_*(x(0))=
\frac{1}{2} x(0)^T \tilde{\Gamma}^d 
(0)x(0)
+\tilde{\beta}^d(0)^{T} x(0)+\tilde{\delta}^d(0),
\end{equation}
where $\tilde{\Gamma}^d$, $\tilde{\beta}^d$ and $\tilde{\delta}^d$ are respectively 
matrix-, vector-, and real-valued functions satisfying the following backward 
propagating differential equations:
\begin{subequations}
\begin{align}
&\dot{\tilde{\Gamma}}^d- 
\tilde{\Gamma}^d \tilde{B}\tilde{R}^{-1}
\tilde{B}^{T}\tilde{\Gamma}^d
+\tilde{\Gamma}^d \tilde{A}+
\tilde{A}^{T}\tilde{\Gamma}^d +\tilde{Q}
=0  \label{eq:abd-1} \\
&\dot{\tilde{\beta}}^d=
\Big( 
\tilde{\Gamma}^d \tilde{B}\tilde{R}^{-1}
\tilde{B}^{T}-\tilde{A}^{T}\Big )
\tilde{\beta}^d \label{eq:abd-2} \\
&\dot{\tilde{\delta}}^d=\frac{1}{2} 
(\tilde{\beta}^d)^{T}\tilde{B}
\tilde{R}^{-1}
\tilde{B}^{T}\tilde{\beta}^d 
\label{eq:abd-3} 
\end{align}
\end{subequations}
with the final conditions $\tilde{\Gamma}^d (T)=\tilde{M}^d$, 
$\tilde{\beta}^d (T)=-
 \tilde{M}^dd$ and 
$\tilde{\delta}^d (T)=\frac{1}{2}
d^T \tilde{M}^dd$.

We summarize the above analysis in the following theorem.
\begin{thm} \label{theo-op}
The social planner problem 
(\ref{eq-social cost}) has an optimal control law $u^{v}_*$ defined
in (\ref{cen-ctr}) and corresponding to
\[
J^{v}_*=\min\limits_{d \in \Delta^N}
 J^{d}_*.
\]
\end{thm}

As discussed in Section \ref{model}, to capture the discrete choice 
phenomenon, the final cost forces the agents to be at time $T$ in the 
vicinity of one of the destination points. Indeed, the following theorem 
establishes that for sufficiently large $M_{ij}$, each player reaches 
an arbitrarily small neighborhood of a destination point. 
Moreover, it asserts that 
there is only one set of destination points
$p^* \in \mathbb{R}^{Nn}$ that the agents can reach exactly under 
an optimal control law, namely, 
the final state $x_0(T)$ under the control law $u_0$ optimizing
\begin{equation} \label{rest}
J_0(u) = \int_0^T \left \{\frac{1}{2} x^T\tilde{Q}x + 
\frac{1}{2} u^T \tilde{R} u \right\} \mathrm{dt},
\end{equation}
i.e., (\ref{partial-cost}) without the final cost.
\begin{thm} \label{reachability}
Suppose that $(A_i,B_i)$, $i=1,\dots,N$, are controllable and the
agents are minimizing (\ref{eq-social cost}). Then, 
\begin{enumerate}[i.]
\item \label{reach-ball} for any $\epsilon>0$, there exists $M_0>0$ such that 
for all $M_{ij}>M_0$, each agent is 
at time $T$ inside a ball of radius 
$\epsilon$ 
and centered at one of the destination points.
\item the agents $1,\dots,N$ reach at time $T$ the destination points 
$d=(d_1,\dots,d_N) \in \Delta^N$ if and only if $d=p^*$.
\end{enumerate}  
\end{thm}
\begin{IEEEproof}
Let $\epsilon>0$ and
$d\in \Delta^N$. The pairs 
$(A_i,B_i)$, for $i=1,\dots,N$, are controllable. Therefore, there 
exist $N$ continuous control laws $\tilde{u}_i$, $i=1,\dots,N$, 
such that the corresponding final states satisfy
$\tilde{x}_i(T)=d_i$, $i=1,\dots,N$. Let 
$\tilde{u}=(\tilde{u}_{i},\tilde{u}_{-i})$. 
We have
\[
J^d(\tilde{u}) = 
 \int_0^T \left \{\frac{1}{2} \tilde{x}^T\tilde{Q}\tilde{x} + 
\frac{1}{2} \tilde{u}^T \tilde{R} \tilde{u} \right\} \mathrm{dt}. 
\] 
By optimality, we have
\[
\sum_{i=1}^N \frac{M_{id_i}}{2}
\left\|x_i(u^{d}_*)(T)-d_i \right\|^2
\leq 
J^{d}_*\leq J^d
(\tilde{u}).
\]
The cost $J^d(\tilde{u})$ is independent of $M_{ij}$. Therefore, there exists
$M^0_{d}>0$ such that for all $M_{id_i}>M^0_{d}$, $\|x_i(u^{d}_*)(T)-d_i\|^2<\epsilon$,
for $i=1,\dots,N$. By choosing $M_0=\max\limits_{d \in \Delta^N} M_{d}^0$, we get \ref{reach-ball}).

Next, suppose that $d \neq p^*$ for all $d \in \Delta^N$. 
The optimal social cost is $J^{d}_*$, for some $d$ and some
$M_{id_i}$, $i=1,\dots,N$. We suppose that the players reach under their optimal strategies 
the destination points $d_1,\dots,d_N$. Let $M'_{id_i}>M_{id_i}$ for $i=1,
\dots,N$.
We have, for all $u \in U^N$, $J'(u)\geq 
J^d(u)$ where
\begin{equation*}
J'(u) = 
 \int_0^T \left \{\frac{1}{2} x^T\tilde{Q}x + 
\frac{1}{2} u^T \tilde{R} u \right\} \mathrm{dt} + 
 \sum_{i=1}^N \frac{M'_{id_i}}{2}
\|x_i(T)-d_i\|^2.
\end{equation*}   
Under $u^{d}_*$, the players $1,\dots,N$ reach $d_1,\dots,d_N$. Therefore, 
\[
J'\left(u^{d}_*\right)=
J^d\left(u^{d}_*\right)=\min_u J^d
(u)=J^{d}_*.
\]
Therefore, 
$\min\limits_{u} J'(u)=\min\limits_u 
J^d (u)$.
This equality holds for all 
$M'_{id_i}>M_{id_i}$, $i=1,\dots,N$.
The solutions of 
(\ref{eq:abd-1})-(\ref{eq:abd-3}) are 
analytic functions
of $\tilde{M}^d$ (for a proof 
of the analyticity one can refer to
\cite{walter1998ordinary}). Therefore,
the optimal cost $\min_{u} J'(u)$ 
defined
in (\ref{cen-cost}) is an analytic
function of $M'_{id_i}$. But $\min_{u} 
J'(u)$ is constant for all  $M'_{id_i}
>M_{id_i}$. Therefore, by analyticity, 
it is constant 
for all $M'_{id_i} \geq 0$, and
more precisely for $M'_{id_i} = 0$. This implies
that $u^{d}_*$ is the optimal
control law of $J_0(u)$ defined in
(\ref{rest}). The definition
of $p^*$ implies that 
$x\left(u^{d}_*\right)(T)=p^* \neq d$. 
This is a contradiction, so in fact some 
of the agents cannot reach their destination point. 

Now suppose that there exists $v \in \Delta^N$ such that $v=p^*$. We have
$J^v(u) \geq J_0(u)$ for all $u$. Following the definition of $p^*$, we have 
\[
\min\limits_uJ_0(u)=J_0(u_0)= J^v(u_0).
\] 
Therefore, the optimal control of $J^v$ is $u^{v}_*=u_0$. Hence, 
the agents reach $p^*=v$. 
\end{IEEEproof}

\begin{rem} \label{initial tendency}
We show in this remark that in the absence 
of a social effect ($q=0$),
the number of agents that go towards a 
destination point $p_j$ decreases as 
$M_{ij}$ increases.  To simplify things, we 
consider the binary choice 
case $l=2$. In the absence of a social
effect, each agent $i$ minimizes its individual cost (\ref{ind_cost_min}). 
In the following, we write
$J_i^{p_j}\left(u_i,x^{(N)}\right)$ as $J_i^{p_j}\left(u_i,M \right)$ to emphasize that the 
coefficient 
$M_{ij}$ in $J_i^{p_j}\left(u_i,x^{(N)}\right)$ is 
equal to $M$, and that the cost does not 
depend on $x^{(N)}$ ($q=0$). Following Theorem \ref{theo-op}
and the absence of a social effect, for 
$M_{i1}=M_1>0$ and $M_{i2}=M_2>0$,
an agent $i$ goes towards $p_1$ if and only 
if 
$\min J_i^{p_1}\left(u_i,M_1\right)<\min J_i^{p_2}\left(u_i,M_2\right)$. 
For an $M'_2>M_2$, 
$\min J_i^{p_1}\left(u_i,M_1\right)<\min J_i^{p_2}\left(u_i,M_2\right) \leq 
\min J_i^{p_2}\left(u_i,M'_2\right)$.
Therefore,
by increasing $M_2$, the number of agents that go towards $p_2$ decreases.
\end{rem} 
 
A naive approach to find an exact social optimum would be to solve the $l^N$ LQR problems
corresponding to the different combinations of destinations. 
This is obviously computationally expensive, and moreover, with this approach 
each player needs to observe the states and parameters of all the other players. 
Instead, we develop in the following sections a set of decentralized strategies that 
are asymptotically optimal. These strategies are decentralized in the sense that an 
agent $i$'s strategy depends only on its state $x_i$ and on the distributions $\mathbb{P}_0$ 
and $\mathbb{P}_\theta$ of the initial conditions and parameters respectively.
\section{Decentralized Social Optimum} \label{decen}

A weaker solution concept than the social optimum is the person-by-person 
optimal solution \cite{ho1972team, yuksel2013stochastic}. 
\begin{definition}
A strategy profile $(u_i^*,u_{-i}^*)$
is said to be person-by-person optimal with respect to 
the social cost $J_{soc}(u_i,u_{-i})$ if
for all $i \in \{1,\dots,N\}$, for all 
$u_i \in U$,
$J_{soc}(u_i,u^*_{-i}) \geq J_{soc}
(u_i^*,u^*_{-i})$. 
\end{definition}
A social optimum is necessarily a person-by-person optimal solution. 
Following the methodology proposed in \cite{huang2012social}, we compute 
in the following section a set of decentralized approximate person-by-person 
solutions. Moreover, we show under some technical assumptions that these 
solutions become socially optimal as $N \to \infty$.

\subsection{Person-by-Person Optimality} \label{reduced-problem}

Assuming that the other players fixed their person-by-person optimal strategies 
$u_{-i}^*$, an agent $i$ computes its person-by-person optimal strategy $u^*_i$ 
by minimizing the cost $J_{soc}(u_i,u^*_{-i})$ over the strategies $u_i \in U$. 
Similarly to \cite{huang2012social}, one can show that the social cost can be written 
\[
J_{soc}(u_i,u_{-i}^*)=J_{1,i}
\left(u_i,x^{*(N)}_{-i} \right)
+J_{2,i}(u_{-i}^*),
\]
where $x^{*(N)}_{-i}=1/N\sum_{j=1,j\neq i}^N x_j^*$,
\iftoggle{jou}{
\begin{align*}
J_{1,i}\left(u_i,x^{*(N)}_{-i}\right)&=
 \int_0^T \left\{ x_i^T\hat{Q}x_i + \left(x^{*(N)}_{-i}
 \right)^T \hat{L} x_i 
+ \frac{r_i}{2} \|u_i\|^2 \right\} \mathrm{dt} \\ 
&+ \min\limits_{j=1,\dots,l}
\frac{M_{ij}}{2}\|x_i(T)-p_j\|^2\\
\hat{Q}&=\frac{q}{2}\left(I_n-\frac{Z}{N} 
\right)^T
\left(I_n-\frac{Z}{N}\right)
+\frac{q(N-1)}{2N^2}Z^TZ \\
\hat{L}&=-qZ^T
\left(I_n-\frac{Z}{N} \right)-qZ +\frac{q(N-1)}{N} Z^TZ. 
\end{align*}}{
\begin{align*}
J_{1,i}\left(u_i,x^{*(N)}_{-i}\right)&=
 \int_0^T \left\{ x_i^T\hat{Q}x_i + \left(x^{*(N)}_{-i}
 \right)^T \hat{L} x_i 
+ \frac{r_i}{2} \|u_i\|^2 \right\} \mathrm{dt}  
+ \min\limits_{j=1,\dots,l}
\frac{M_{ij}}{2}\|x_i(T)-p_j\|^2\\
\hat{Q}&=\frac{q}{2}\left(I_n-\frac{Z}{N} 
\right)^T
\left(I_n-\frac{Z}{N}\right)
+\frac{q(N-1)}{2N^2}Z^TZ \\
\hat{L}&=-qZ^T
\left(I_n-\frac{Z}{N} \right)-qZ +\frac{q(N-1)}{N} Z^TZ. 
\end{align*}
}
The term $J_{2,i}(u_{-i}^*)$ does not depend on the strategy $u_i$
of player $i$. Therefore, minimizing $J_{soc}(u_i,u^*_{-i})$ reduces to 
minimizing $J_{1,i}\left(u_i,x^{*(N)}_{-i}\right)$.

The person-by-person optimal solutions $(u_i^*,u_{-i}^*)$ are fixed points of 
the following system of equations:
\begin{align*}
u_i^*= 
\argmin_{u_i \in U} J_{1,i} \left(u_i,
{x^*_{-i}}^{(N)} \right) && i=1,\dots,N.
\end{align*}  
Equivalently, these solutions are  the Nash equilibria of a noncooperative game 
involving the $N$ players defined in (\ref{eq-dynamics}) but associated with 
the individual costs
\begin{align} \label{noncoop}
J_{1,i}\left(u_i,{x_{-i}}^{(N)}
\right) && 
i=1,\dots,N.
\end{align} 
The players are cost coupled through the average of the population. In the 
following we develop via the MFG approach a decentralized approximate 
Nash strategy profile with respect to (\ref{noncoop}), or equivalently 
a set of decentralized approximately person-by-person optimal strategies 
with respect to (\ref{eq-social cost}).
\subsection{Mean Field Equation System}

According to the MFG approach, each agent assumes a continuum of agents and 
computes its best response to an assumed given continuous path $\bar{x}$. 
This path represents the mean path of the infinite size population under 
a Nash strategy profile. Since the players must collectively reproduce 
this assumed mean path when applying their best responses to it, this path 
can be computed by a fixed point argument. Under the infinite size 
population assumption, the costs (\ref{noncoop}) reduce to the cost of a generic agent with state
$x$, control input $u$ and parameters $\theta$:
\iftoggle{jou}{
\begin{multline} \label{aux-cost}
J(u,\bar{x},x^0,\theta)=
\int_0^T \bigg\{ \frac{q}{2}
\|x\|^2 + q \bar{x}^{T} L x 
+ \frac{r_\theta}{2} \|u\|^2 \bigg\} \mathrm{dt} 
\\ 
+ \min\limits_{j=1,\dots,l}
\bigg\{\frac{M_{\theta j}}{2}\|
x(T)-p_j\|^2\bigg\}, 
\end{multline}}{
\begin{equation} \label{aux-cost}
J(u,\bar{x},x^0,\theta)=
\int_0^T \bigg\{ \frac{q}{2}
\|x\|^2 + q \bar{x}^{T} L x 
+ \frac{r_\theta}{2} \|u\|^2 \bigg\} \mathrm{dt}  
+ \min\limits_{j=1,\dots,l}
\bigg\{\frac{M_{\theta j}}{2}\|
x(T)-p_j\|^2\bigg\}, 
\end{equation}
}
where $\bar{x}=\mathbb{E}x$ is the mean trajectory of the 
infinite size population. The 
generic agent's state $x$ satisfies (\ref{eq-dynamics}) where $(A_i,B_i,u_i)$ is 
replaced by $(A_\theta,B_\theta,u)$, with an initial state $x^0(\omega)$ drawn from
$\mathbb{P}_0$ and parameters 
$\theta(\omega)=(A_\theta,B_\theta,r_\theta,M_{\theta 1},\dots,M_{\theta l})(\omega)$
drawn from $\mathbb{P}_\theta$. 
In the following, we omit $\omega$ from the notation.

\subsubsection{The Generic Agent's Best Response to $\bar{x}$}

We define 
$\Gamma_k^{\theta} \in C([0,T],\mathbb{R}^{n \times n})$, 
$\beta_k^{\theta} \in C([0,T],\mathbb{R}^n)$ and $\delta_k^{\theta} \in C([0,T],\mathbb{R})$
to be the unique solutions of the following backward propagating differential equations:
\begin{subequations}
\begin{align}
&\dot{\Gamma}_k^{\theta}-\frac{1}{r_\theta} 
\Gamma^{\theta}_k B_\theta B_\theta^{T}\Gamma_k^\theta 
+\Gamma_k^\theta A_\theta +A_\theta^{T}\Gamma_k^\theta 
+q I_{n}=0  \label{eq:gamma-1-n} \\
&\dot{\beta}_k^{\theta}=\left(\frac{1}{r_\theta} 
\Gamma_k^\theta B_\theta B_\theta^{T}-A_\theta^{T}\right )
\beta_k^{\theta} -qL
\bar{x} \label{eq:beta-2-n} \\
&\dot{\delta}_k^{\theta}=\frac{1}{2r_\theta} 
(\beta_k^{\theta})^{T} B_\theta B_\theta^{T}\beta_k^{\theta},  \label{eq:delta-3-n} 
\end{align}
\end{subequations}
with the final conditions 
\[
\Gamma_k^\theta (T)=M_{\theta k}I_{n}, \;\; \beta_k^{\theta} (T)=-M_{\theta k}p_k, 
\;\; \delta_k^{\theta} (T)=\frac{1}{2} M_{\theta k} p_k^{T} p_k.\]
\begin{lem} \label{lemma:tracking}
Given the initial condition and the parameters, an agent's best response 
to $\bar{x}$ and the corresponding optimal cost are
\iftoggle{jou}{
\begin{align}\label{eq:ctr}
&\hat{u}\left(t,x^0,\theta\right) =\nonumber\\
&\sum_{j=1}^l - \frac{1}{r_\theta} B_\theta^T 
\left (\Gamma_j^\theta(t) \hat{x}
\left(t,x^0,\theta\right) + \beta_j^\theta(t) \right ) 
\mathbbm{1}_{D_j^\theta(\bar{x})}(x^0)  
\end{align}}{
\begin{equation}\label{eq:ctr}
\hat{u}\left(t,x^0,\theta\right) =
\sum_{j=1}^l - \frac{1}{r_\theta} B_\theta^T 
\left (\Gamma_j^\theta(t) \hat{x}
\left(t,x^0,\theta\right) + \beta_j^\theta(t) \right ) 
\mathbbm{1}_{D_j^\theta(\bar{x})}(x^0)  
\end{equation}
}
\iftoggle{jou}{
\begin{align} \label{eq:cost:opt}
 & J^*\left(\bar{x},x^0,\theta\right) =
 \nonumber \\
 & \sum_{j=1}^l
 \left (\frac{1}{2}(x^0)^T\Gamma_j^\theta(0)x^0 + 
 (\beta^\theta_j)(0)^Tx^0+ \delta_j^\theta(0) \right )\mathbbm{1}_{D_j^\theta(\bar{x})}(x^0),
\end{align}}{
\begin{equation} \label{eq:cost:opt}
  J^*\left(\bar{x},x^0,\theta\right) =
  \sum_{j=1}^l
 \left (\frac{1}{2}(x^0)^T\Gamma_j^\theta(0)x^0 + 
 (\beta^\theta_j)(0)^Tx^0+ \delta_j^\theta(0) \right )\mathbbm{1}_{D_j^\theta(\bar{x})}(x^0),
\end{equation}
}
where $ \hat{x}\left(t,x^0,\theta\right)$ is the generic agent's state under the 
feedback law (\ref{eq:ctr}), $\Gamma_k^\theta$, $\beta^\theta_k$, $\delta_k^\theta$ 
are the unique solutions of (\ref{eq:gamma-1-n})-(\ref{eq:delta-3-n}), and
\begin{multline} \label{eq:da}
D_j^\theta(\bar{x}) =  \bigg \{x\in \mathbb R^n \bigg|  \, \forall k=1,\dots,l, \,\,
\frac{1}{2}x^T \Big (\Gamma^\theta_{j}(0)-\Gamma^\theta_{k}(0) \Big ) x+\\  
 \Big(\beta^\theta_{j}(0)-\beta^\theta_{k}(0)\Big)^{T} x +
 \delta^\theta_{j}(0)-\delta^\theta_{k}(0) \leq 0 \bigg \}. 
\end{multline}
\end{lem}
\begin{IEEEproof}
See \cite[Lemma 1]{DBLP:journals/corr/SalhabMN15}.
\end{IEEEproof}
The cost function (\ref{aux-cost}) can be written as the minimum of $l$ 
LQR cost functions each corresponding to a 
distinct possible destination point. When 
minimizing one of these LQR costs, an agent goes towards the corresponding
destination point. The region $D_j^\theta(\bar{x})$ defined in \eqref{eq:da} 
includes the initial conditions for which the LQR problem corresponding to $p_j$ 
is the less costly LQR problem. Therefore, there exist $l$ basins of 
attraction $D_j^\theta(\bar{x})$, $j=1,\dots,l$, where the players initially 
present in $D_j^\theta(\bar{x})$ go towards $p_j$.


We define 
$\Psi_j^\theta(\eta_1,\eta_2,\eta_3,\eta_4)=\Phi_j^\theta(\eta_1,\eta_2)^TB_\theta B_\theta^{T}\Phi_j^\theta(\eta_3,\eta_4)$, 
where $\Pi_{j}^{\theta}(t)=\frac{1}{r_\theta}\Gamma_j^{\theta}(t)B_\theta B_\theta^T-A_\theta^T$ 
and $\Phi_j^\theta$ is the unique solution of 
\begin{equation} \label{TM}
\frac{d\Phi_j^\theta(t,\eta)}{dt}=
\Pi_j^\theta(t)\Phi_j^\theta(t,\eta),\,\,\,\, 
\Phi_j^\theta (\eta,\eta)=I_n .
\end{equation}

The state trajectory of the generic agent is then \cite{DBLP:journals/corr/SalhabMN15}
\iftoggle{jou}{
\begin{multline} \label{state-opt} 
\hat{x}
\left(t,x^0,\theta\right)  =\sum_{j=1}^{l}  
\mathbbm{1}_{D^{\theta}_{j}(\bar{x})}
(x^0) \Big \{\Phi_j^\theta
(0,t)^Tx^0 \\
+\frac{M_{\theta j}}{r_\theta}\int_0^t \! \ \Psi_j^{\theta}
(\sigma,t,\sigma,T)p_{j} \, \mathrm{d}\sigma  \\
 +\frac{q}{r_\theta}\int_0^t \! \ \int_T^\sigma \! \ 
 \Psi_j^{\theta}(\sigma,t,\sigma,\tau)
 L\bar{x}(\tau) \, \mathrm{d}\tau 
 \mathrm{d}\sigma \Big \}. 
\end{multline}}{
\begin{multline} \label{state-opt} 
\hat{x}
\left(t,x^0,\theta\right)  =\sum_{j=1}^{l}  
\mathbbm{1}_{D^{\theta}_{j}(\bar{x})}
(x^0) \Big \{\Phi_j^\theta
(0,t)^Tx^0
+\frac{M_{\theta j}}{r_\theta}\int_0^t \! \ \Psi_j^{\theta}
(\sigma,t,\sigma,T)p_{j} \, \mathrm{d}\sigma  \\
 +\frac{q}{r_\theta}\int_0^t \! \ \int_T^\sigma \! \ 
 \Psi_j^{\theta}(\sigma,t,\sigma,\tau)
 L\bar{x}(\tau) \, \mathrm{d}\tau 
 \mathrm{d}\sigma \Big \}. 
\end{multline}
}
\subsubsection{Existence of a Solution for the Mean Field Fixed Point Equation System} \label{exist_fix}

The mean field equation system
is determined by 
 \eqref{eq:gamma-1-n}-\eqref{eq:delta-3-n} 
plus the infinite size population mean equation
\begin{equation} \label{additional-MFE}
\bar{x}(t)=\int  \hat{x}
\left(t,x^0,\theta\right) \mathrm{d}\mathbb{P}_0 \times \mathbb{P}_\theta.
\end{equation}
This equation system defines an operator $G(.)$
from the Banach space 
$(C([0,T],\mathbb{R}^n),\|\|_\infty)$ into
itself. In fact, given a continuous path
$\bar{x}$, one can solve \eqref{eq:gamma-1-n}-\eqref{eq:delta-3-n} and compute by (\ref{additional-MFE}) 
the mean
trajectory $G(\bar{x})$ of the generic
agent when it optimally
tracks $\bar{x}$.
We define 
\begin{equation} \label{bounds}
\begin{split}
k_1&=\mathbb{E}\|x^0\|\times\left(\sum_{j=1}^l \max\limits_{(\theta,t)\in \Theta \times 
[0,T]}\|\Phi_j^\theta(0,t)\|\right)\\
k_2&=\sum_{j=1}^l\max\limits_{(\theta,t)\in \Theta \times [0,T]}
\bigg \|\frac{M_{\theta j}}{r_\theta}\int_0^t \! \ \Psi_j^{\theta}(\sigma,t,\sigma,T)p_{j} \, \mathrm{d}\sigma \bigg \| \\ 
k_3&=\sum_{j=1}^l\max\limits_{(\theta,t,\sigma,\tau)\in \Theta \times 
[0,T]^3}\frac{q}{r_\theta}\|\Psi_j^{\theta}(\sigma,t,\sigma,\tau)L\|.
\end{split}
\end{equation}
Since $\Theta$ and $[0,T]$ are compact and $\Phi_j^\theta$ is continuous
with respect to time and parameter $\theta$, then $k_1$, $k_2$ and $k_3$ are 
well defined.
\begin{assum}	\label{assumption: bounds}
We assume that $\sqrt{\max(k_1+k_2,k_3)}T<\pi/2$. 
\end{assum}
Noting that the left hand side of the inequality tends to zero as $T$ goes to zero, Assumption \ref{assumption: bounds} can be satisfied for short time horizon $T$ for example.
\begin{assum} \label{assum-neg}
We assume that $L \succeq 0$, where $L$
is defined in (\ref{L-eq}).
\end{assum}
Assumption \ref{assum-neg} is satisfied, 
for example, when $Z=-\alpha I_n$, $\alpha>0$. In this case, the 
social effect pushes the agents away from the
mean of the population.
\begin{assum} \label{assum-P0}
We assume that $\mathbb{P}_0$
is such that the $\mathbb{P}_0-$measure of quadric surfaces is zero.
\end{assum}  
\begin{assum} \label{assum-Ex2}
We assume that $\mathbb{E}\|x^0\|^2<\infty$.
\end{assum}

\begin{thm} \label{lemma:existanceinit}
Under Assumptions \ref{assumption: bounds}, \ref{assum-P0} and \ref{assum-Ex2}, 
$G$ has a fixed point. If $(A_\theta,B_\theta,M_{\theta j},r_\theta)=(A,B,M_{j},r)$, i.e., the
parameters are the same for all the agents, the result holds with Assumption 
\ref{assumption: bounds} replaced by Assumption \ref{assum-neg}.
\end{thm}

\begin{IEEEproof}
See \cite[Theorems 6 and 8]{DBLP:journals/corr/SalhabMN15}.
\end{IEEEproof}

Theorem \ref{lemma:existanceinit} provides conditions under which
a solution of the mean field equations
\eqref{eq:gamma-1-n}-\eqref{eq:delta-3-n} and \eqref{additional-MFE}
exists.
In case of nonuniform parameters, i.e. 
$(A_\theta,B_\theta,M_{\theta j},r_\theta)$ are not the same for all the agents, the existence 
of a fixed point is proved by Schauder's fixed point theorem
\cite[Theorem 8]{DBLP:journals/corr/SalhabMN15}
, where Assumption
\ref{assumption: bounds} is used to construct
a bounded set that is mapped by $G$ into 
itself. When the parameters are the same for 
all the agents,  by similar techniques than
those used in\cite[Theorem 6]{DBLP:journals/corr/SalhabMN15}, one can show
that
a fixed point of $G$ is the 
optimal state of an LQR problem of running cost $\frac{q}{2}x^T(L+I_n)x+\frac{r}{2}\|u\|^2$. The existence and uniqueness 
of an optimal solution of this LQR problem
is a consequence of Assumption 
\ref{assum-neg}.
In the following, \eqref{eq:ctr} and \eqref{state-opt} are considered for a fixed
point path $\bar{x}$. We define
\begin{equation} \label{finit_m}
\hat{x}^{(N)}
(t)=\frac{1}{N}\sum_{i=1}^N\hat{x}_i(t)=\int\hat{x}\left(t,x^0_i,\theta_i\right)
\mathrm{d}
\mathbb{P}_0^N(x_i^0)
\mathrm{d}\mathbb{P}_\theta^N(\theta_i),
\end{equation}
and $\hat{u}^{(N)}=(\hat{u}_i,\hat{u}_{-i})$,
where $\hat{u}_i(t)=\hat{u}(t,x^0_i,\theta_i)$ and 
$\hat{x}_i(t)=\hat{x}(t,x^0_i,\theta_i)$.
\subsection{Asymptotic Social Optimum}

In this section, we show that when the agents 
apply the strategy profile $\hat{u}$ defined below Theorem \ref{lemma:existanceinit} and in (\ref{eq:ctr}), 
the corresponding per agent social cost (\ref{eq-social cost}) converges to the optimal per agent social cost  
as the size of the population increases to infinity. 
At the end of this section, we also give an explicit form of the asymptotic per agent optimal social cost.
\begin{assum}	\label{assumption: finite mean 2}
We assume that $\frac{1}{N}\sum_{i=1}^N\|x^0_i\|^2 
<
C$ for all
$N>0$.
\end{assum}
\begin{rem} \label{assum}
Assumption \ref{assumption: finite mean 2} implies Assumption \ref{assum-Ex2}. 
In fact, $\mathbb{P}_0^{N}$ converges in distribution to $\mathbb{P}_0$. Therefore, there 
exists on some probability space $(\Omega,\mathcal{F},\mathbb{P})$ a sequence of random variables $X^0_N$ of distribution $\mathbb{P}_0^N$ and 
a random variable $X^0$ of distribution $\mathbb{P}_0$ such that $X_N^0$ converges
with probability one to $X^0$.
We may consider, without loss of generality,
that $(\Omega,\mathcal{F},\mathbb{P})$ is 
the same as the one defined in Section \ref{model}.
By Fatou's Lemma \cite{Rudin87_Real}, 
\begin{multline*}
C \geq \liminf_{N} \frac{1}{N}\sum_{i=1}^N
\left\|
x^0_i \right\|^2 = \liminf_{N} \int \left\|X_N^0\right\|^2 \mathrm{d}\mathbb{P} \\ 
\geq \int \liminf_{N} \left \|X_N^0
\right\|^2 \mathrm{d} \mathbb{P} 
= \int \left \|X^0 \right\|^2 \mathrm{d} \mathbb{P} = \mathbb{E} 
\left\| x^0  \right\|^2.
\end{multline*} 
\end{rem}
The functions defined by (\ref{eq:gamma-1-n}), (\ref{eq:beta-2-n}) and (\ref{eq:delta-3-n}) are 
continuous with respect to $\theta$, which belongs to a compact set $\Theta$. The random
variables
$\theta$ and $x^0$ are assumed to be independent. 
Therefore, under Assumption \ref{assum-Ex2} and
by Fubini-Tonelli's theorem \cite{Rudin87_Real}, the operator $G$ defined in paragraph \ref{exist_fix}  by \eqref{eq:gamma-1-n}-\eqref{eq:delta-3-n} and  (\ref{additional-MFE}) has the following form:

\begin{equation}  \label{meaninit}
\mathbb{E}\hat{x}\left(t,x^0,\theta
\right)
= G (\bar{x})(t)
= \int_\Theta \, 
\int_{\mathbb{R}^n} \! \hat{x}\left(t,x^0,\theta\right) \mathrm{d}
\mathbb{P}_0(x^0)\mathrm{d}\mathbb{P}_\theta(\theta). 
\end{equation}

In the following lemma, we show that when applying the decentralized 
person-by-person control laws, the finite population average path converges 
to the fixed point path $\bar{x}$ that the agents are optimally tracking. 
In the standard LQG MFG literature, the proof of this result relies on the uniform 
boundedness and equicontinuity of the generic agent's state trajectory with respect 
to the initial conditions and parameters. 
In our case, this trajectory (\ref{state-opt}), considered as a function
of the time $t$, the 
initial condition $x^0$
and the 
parameter $\theta$, is discontinuous. In fact, it has on each basin of attraction 
$D_j^\theta$ a different structure that
depends on the corresponding $p_j$.
Hence, the proof requires some additional constructions to deal with the discontinuity.

\begin{lem} \label{appr}
Under Assumptions \ref{assum-P0} and \ref{assumption: finite mean 2},
\begin{equation} \label{conv_ave}
\lim\limits_{N \rightarrow \infty } 
\int_0^T \left \|\hat{x}^{(N)}-\bar{x} \right\|^2 \mathrm{dt}=0.
\end{equation}
\end{lem}
\begin{IEEEproof}
In view of (\ref{finit_m}) and (\ref{meaninit}), we have
\iftoggle{jou}{
\begin{align*}
\hat{x}^{(N)}(t)-\bar{x}(t)&=\int\hat{x}\left(t,x^0_i,\theta_i\right)
\mathrm{d}
\mathbb{P}_0^N(x_i^0)
\mathrm{d}\mathbb{P}_\theta^N(\theta_i)\\
&-
\int\hat{x}\left(t,x^0,\theta\right)
\mathrm{d}
\mathbb{P}_0(x^0)
\mathrm{d}\mathbb{P}_\theta(\theta).
\end{align*}}{
\begin{equation*}
\hat{x}^{(N)}(t)-\bar{x}(t)=\int\hat{x}\left(t,x^0_i,\theta_i\right)
\mathrm{d}
\mathbb{P}_0^N(x_i^0)
\mathrm{d}\mathbb{P}_\theta^N(\theta_i)-
\int\hat{x}\left(t,x^0,\theta\right)
\mathrm{d}
\mathbb{P}_0(x^0)
\mathrm{d}\mathbb{P}_\theta(\theta).
\end{equation*}
}
If $\hat{x}(t,x_i^0,\theta_i)$ and 
$\hat{x}(t,x^0,\theta)$ were uniformly 
bounded and equicontinuous with respect to
the initial conditions and parameters, 
then one can show the convergence  
by \cite[Corollary 1.1.5]{stroock1979multidimensional}. But 
$\hat{x}(t,x_i^0,\theta_i)$ and 
$\hat{x}(t,x^0,\theta)$ are discontinuous. 
Alternatively, we show 
that the set of discontinuity points has 
a measure zero under Assumption \ref{assum-P0}. We then 
we show that $\hat{x}^{(N)}$ converges pointwise to $\bar{x}$. Finally, 
We prove the uniform convergence, from which the result follows.

\emph{Pointwise convergence.}
$\mathbb{P}_0^N$ and $\mathbb{P}_\theta^N$ converge respectively in 
distribution to $\mathbb{P}_0$ and $\mathbb{P}_\theta$. Therefore, there exist
on some probability space $(\Omega,\mathcal{F},\mathbb{P})$
a sequence of random variables $X^0_N$ of distribution $\mathbb{P}_0^N$ 
(resp. a sequence of random variables $\xi^\theta_N$ of distribution $\mathbb{P}_\theta^N$), 
and a random variable $X^0$ of distribution $\mathbb{P}_0$ (resp. a random variable
$\xi^\theta$ of distribution $\mathbb{P}_\theta$) such that $X_N^0$ (resp. $\xi_N^\theta$) 
converges with probability one to $X^0$ (resp. $\xi^\theta$). Thus,
\begin{equation*}
\hat{x}^{(N)}(t)-\bar{x}(t)
=\int
\left(\hat{x}\left(t,X^0_N ,\xi_N^\theta\right) -  
\hat{x}\left(t,X^0 ,\xi^\theta\right)\right)\mathrm{d}\mathbb{P}.
\end{equation*}
For a fixed $t$, the discontinuity points of $\hat{x}\left(t,x^0,\theta\right)$
(considered now as a function of $x^0$ and $\theta$)
are included in the set 
$D=\{(x^0,\theta) \in \mathbb{R}^n \times \Theta\, 
| \, x^0 \in \partial D_j^\theta (\bar{x})\}$. 
Under 
Assumption \ref{assum-P0} and the independence 
of $x^0$ and $\theta$, one can prove that $
\mathbb{P}_0\times \mathbb{P}_\theta(D)=0$. 
Hence, $\hat{x}\left(t,X^0_N ,\xi_N^\theta\right)$ converges with 
probability one
to $\hat{x}\left(t,X^0 ,\xi^\theta\right)$. The compactness of 
$[0,T]$ 
and $\Theta$, and the continuity of $\Pi_j^\theta$ 
imply 
\[\left\|\hat{x}\left(t,X^0_N ,\xi_N^\theta\right) -  
\hat{x}\left(t,X^0 ,\xi^\theta\right)\right\| \leq
 K_1 \|X^0_N\| + K_2 \|X^0\| + K_3,\] for some finite $K_1, 
 K_2,K_3>0$. $\hat{x}^{(N)}(t)$ converges pointwise 
to $\bar{x}(t)$ for all $t \in [0,T]$ as a consequence of Assumption \ref{assumption: finite mean 2}, Remark \ref{assum} 
and Lebesgue's dominated convergence theorem.

\emph{Uniform convergence.}
As in the proof of Theorem \ref{lemma:existanceinit}, 
see \cite[Theorem 8]{DBLP:journals/corr/SalhabMN15}, one can show that 
for all $t_1, t_2$,
$\left\|\hat{x}^{(N)}(t_1)-\hat{x}^{(N)}(t_2)
\right\| \leq K |t_1-t_2|$ and 
$\left\|\bar{x}(t_1)-\bar{x}(t_2)\right\| \leq K |t_1-t_2| $, 
where
$K>0$ is independent of $N$. We fix an $\epsilon>0$ and  consider a partition $0=t_0<t_1<\dots<t_j=T$ of
$[0,T]$ such that
for all $t,t' \in [t_k,t_{k+1}]$, 
for $N\geq 1$, 
$\|\hat{x}^{(N)}(t)-\hat{x}^{(N)}(t')
\|<
\epsilon$ and $\|\bar{x}(t)-\bar{x}(t')
\|<
\epsilon$ 
By the pointwise convergence, there exists 
$N_0$ such that for all 
$N>N_0$, for $k=1,
\dots,j$, $\|\hat{x}^{(N)}(t_k)-\bar{x}(t_k)\|<\epsilon$. We fix $N>N_0$. For  an arbitrary $t\in [0,T]$, there exists $k$ 
such that $t \in [t_k,t_{k+1}]$. We have 
\iftoggle{jou}{
\begin{align*}
\|\hat{x}^{(N)}(t)-\bar{x}(t)\| \leq& \|\hat{x}^{(N)}(t)-\hat{x}^{(N)}(t_k)\| + \\
&\|\hat{x}^{(N)}(t_k)-\bar{x}(t_k)\|+\|\bar{x}(t_k)-\bar{x}(t)\| \leq 
3\epsilon.
\end{align*} }{
\begin{equation*}
\|\hat{x}^{(N)}(t)-\bar{x}(t)\| \leq \|\hat{x}^{(N)}(t)-\hat{x}^{(N)}(t_k)\| + \|\hat{x}^{(N)}(t_k)-\bar{x}(t_k)\|+\|\bar{x}(t_k)-\bar{x}(t)\| \leq 
3\epsilon.
\end{equation*}
}
This inequality holds for an arbitrary 
$t\in [0,T]$, therefore, $\lim\limits_{N \rightarrow \infty}
\sup\limits_{t\in [0,T]}\|\hat{x}^{(N)}(t)-\bar{x}(t)\|^2=0$. 
This implies (\ref{conv_ave}).
\end{IEEEproof}

We now state the main result of this paper, which asserts that under appropriate conditions, 
when the agents apply the mean field person to person optimization based decentralized 
strategies (\ref{eq:ctr}), the per agent social cost converges to the per agent optimal 
social cost as the size of the population increases to infinity. 
To compute its control strategy (\ref{eq:ctr}), each agent only needs to know its 
initial condition, current state, the distributions $\mathbb{P}_0$ and $\mathbb{P}_\theta$ and
a fixed point path $\bar{x}$ of the operator $G$ defined in \eqref{meaninit}.
\begin{thm} \label{th-optim}
Under Assumptions \ref{assum-neg}, \ref{assum-P0}, and \ref{assumption: finite mean 2},
\begin{equation} \label{conv_th}
\lim\limits_{N \rightarrow \infty }\left| \inf\limits_{u \in U^N} 
\frac{1}{N}J_{soc}\left(u,x^{(N)}
\right) - 
\frac{1}{N}J_{soc}\left(\hat{u}^{(N)},\hat{x}^{(N)}
\right) 
\right| 
=0.
\end{equation}
\end{thm}
\begin{IEEEproof}
Let $u \in U^N$ such that 
$J_{soc}\left(u,x^{(N)}\right) \leq 
J_{soc}\left(\hat{u}^{(N)},\hat{x}^{(N)}\right)$. 
Noting (\ref{state-opt}), the compactness of $\Theta$, the continuity of $\Pi_j^\theta(t)$
with respect to $t$ and $\theta$ and Assumption \ref{assumption: finite mean 2}, one can 
prove that $(1/N)J_{soc}\left(\hat{u}^{(N)},\hat{x}^{(N)}
\right)<c_0$,
where $c_0$ is independent of $N$. 
Therefore, $(1/N)J_{soc}\left(u,x^{(N)}\right) < c_0$ and 
\[
\frac{1}{N}\sum_{i=1}^N\int_0^T \Big\{\|u_i\|^2+ 
\|\hat{u}_i\|^2
+ \|x_i\|^2 + \|\hat{x}_i\|^2 \Big \} 
\mathrm{dt} <c_1,
\]
where $c_1>0$ is independent of $N$.
Let $\tilde{x}_i=x_i-\hat{x}_i$ and $\tilde{u}_i=u_i-\hat{u}_i$. We have (\ref{decomp}) below 
\iftoggle{jou}{
\begin{align} \label{decomp}
&\frac{1}{N}J_{soc}\left(u,x^{(N)}\right)=
\frac{1}{N}J_{soc}\left(\hat{u}^{(N)},\hat{x}^{(N)}\right)+
\frac{1}{N}\sum_{i=1}^N \int_0^T r_i
\tilde{u}_i^T 
\hat{u}_i \mathrm{dt} \nonumber\\
 &+ \frac{1}{N}\sum_{i=1}^N  
 \int_0^T \left \{\frac{q}{2}
\left\| \tilde{x}_i - Z \tilde{x}^{(N)}
\right\|^2 
+ \frac{r_i}{2}
\| \tilde{u}_i\|^2 \right \}\mathrm{dt}
\nonumber\\
&+ \frac{1}{N}\sum_{i=1}^N \int_0^T q
\left( \tilde{x}_i - Z \tilde{x}^{(N)}
\right)^T \left( 
\hat{x}_i - Z \hat{x}^{(N)}\right)\mathrm{dt} \\
&+ \frac{1}{N}\sum_{i=1}^N 
\min\limits_{j=1,\dots,l}
\frac{M_{ij}}{2}\|x_i(T)-p_j\|^2 \nonumber\\
&- \frac{1}{N}\sum_{i=1}^N  
\min\limits_{j=1,\dots,l}\frac{M_{ij}}{2}
\|\hat{x}_i(T)-p_j\|^2 .\nonumber
\end{align}}{
\begin{align} \label{decomp}
&\frac{1}{N}J_{soc}\left(u,x^{(N)}\right)=
\frac{1}{N}J_{soc}\left(\hat{u}^{(N)},\hat{x}^{(N)}\right)+
\frac{1}{N}\sum_{i=1}^N \int_0^T r_i
\tilde{u}_i^T 
\hat{u}_i \mathrm{dt} \nonumber\\
 &+ \frac{1}{N}\sum_{i=1}^N  
 \int_0^T \left \{\frac{q}{2}
\left\| \tilde{x}_i - Z \tilde{x}^{(N)}
\right\|^2 
+ \frac{r_i}{2}
\| \tilde{u}_i\|^2 
+  q
\left( \tilde{x}_i - Z \tilde{x}^{(N)}
\right)^T \left( 
\hat{x}_i - Z \hat{x}^{(N)}\right)\right \}\mathrm{dt} \\
&+ \frac{1}{N}\sum_{i=1}^N 
\min\limits_{j=1,\dots,l}
\frac{M_{ij}}{2}\|x_i(T)-p_j\|^2 
- \frac{1}{N}\sum_{i=1}^N  
\min\limits_{j=1,\dots,l}\frac{M_{ij}}{2}
\|\hat{x}_i(T)-p_j\|^2 .\nonumber
\end{align}
}
For a fixed point $\bar{x}$ of $G$, and recalling (\ref{aux-cost}) we have
\iftoggle{jou}{
\begin{align} \label{decomp1}
&J (u_i,\bar{x},x^0_i,\theta_i)=
J(\hat{u}_i,\bar{x},x^0_i,\theta_i)
+\int_0^T \bigg\{ \frac{q}{2}
\|\tilde{x}_i\|^2 
+ \frac{r_i}{2} \|\tilde{u}_i\|^2 \bigg\} 
\mathrm{dt} \nonumber\\
&+\int_0^T q 
\bar{x}^{T} 
L \tilde{x}_i \mathrm{dt}
+\int_0^T \bigg\{ q
\tilde{x}_i^T\hat{x}_i 
+ r_i \tilde{u}_i^T \hat{u}_i \bigg\} 
\mathrm{dt}\\ 
&+ 
\min\limits_{j=1,\dots,l}\frac{M_{ij}}{2}
\|x_i(T)-p_j\|^2- 
\min\limits_{j=1,\dots,l}\frac{M_{ij}}{2}
\|\hat{x}_i(T)-p_j\|^2,\nonumber
\end{align}}{
\begin{align} \label{decomp1}
&J (u_i,\bar{x},x^0_i,\theta_i)=
J(\hat{u}_i,\bar{x},x^0_i,\theta_i)
+\int_0^T \bigg\{ \frac{q}{2}
\|\tilde{x}_i\|^2 
+ \frac{r_i}{2} \|\tilde{u}_i\|^2  
 +q 
\bar{x}^{T} 
L \tilde{x}_i 
+ q
\tilde{x}_i^T\hat{x}_i 
+ r_i \tilde{u}_i^T \hat{u}_i \bigg\} 
\mathrm{dt}\nonumber\\ 
&+ 
\min\limits_{j=1,\dots,l}\frac{M_{ij}}{2}
\|x_i(T)-p_j\|^2- 
\min\limits_{j=1,\dots,l}\frac{M_{ij}}{2}
\|\hat{x}_i(T)-p_j\|^2,
\end{align}
}
Now (\ref{decomp}) and (\ref{decomp1}) yield
\iftoggle{jou}{
\begin{align} \label{decomp2}
&\frac{1}{N}J_{soc}\left(u,x^{(N)}\right)=
\frac{1}{N}J_{soc}\left(\hat{u}^{(N)},\hat{x}^{(N)}
\right)\nonumber\\
&+ \frac{1}{N} \sum_{i=1}^N \left(J(u_i,\bar{x},x^0_i,\theta_i)
- J(\hat{u}_i,\bar{x},x^0_i,\theta_i)\right)\\
&+q \int_0^T  \left(\tilde{x}^{(N)}\right)^T L 
\tilde{x}^{(N)} \mathrm{dt}
 +q  \int_0^T \left(\hat{x}^{(N)}
-\bar{x}\right)^T L \tilde{x}^{(N)}
\mathrm{dt}. \nonumber 
\end{align}}{
\begin{align} \label{decomp2}
&\frac{1}{N}J_{soc}\left(u,x^{(N)}\right)=
\frac{1}{N}J_{soc}\left(\hat{u}^{(N)},\hat{x}^{(N)}
\right)
+ \frac{1}{N} \sum_{i=1}^N \left(J(u_i,\bar{x},x^0_i,\theta_i)
- J(\hat{u}_i,\bar{x},x^0_i,\theta_i)\right)\\
&+q \int_0^T  \left(\tilde{x}^{(N)}\right)^T L 
\tilde{x}^{(N)} \mathrm{dt}
 +q  \int_0^T \left(\hat{x}^{(N)}
-\bar{x}\right)^T L \tilde{x}^{(N)}
\mathrm{dt}. \nonumber 
\end{align}
}
By the bounds $c_0$ and $c_1$, the Cauchy-Schwarz inequality and 
Lemma \ref{appr}, we deduce that
$\epsilon_N = q \int_0^T (\hat{x}^{(N)}-\bar{x})^T L \tilde{x}^{(N)} \mathrm{dt}$
converges to $0$ as $N$ goes to infinity.
The optimization of $\hat{u}_i$ with respect to $J$ and 
Assumption \ref{assum-neg} imply 
$\frac{1}{N}J_{soc}(u,x^{(N)})\geq\frac{1}{N}
J_{soc}(\hat{u}^{(N)},\hat{x}^{(N)})+\epsilon_N$.
\end{IEEEproof}
\begin{rem}[Need for Assumption \ref{assum-neg}] \label{need_L_p}
 In static games, a sufficient condition of the person-by-person solution to
be a social optimum is the convexity and 
smoothness of the costs \cite[Lemma 2.6.1]{yuksel2013stochastic}. 
Although not explicitly mentioned by the authors \cite{huang2012social},
this condition (which is automatically satisfied in the LQG setting) guarantees also the convergence of the person-by-person solution 
to the social optimum in case of dynamic LQG MFG problems \cite[Theorem 4.2]{huang2012social}. 
In fact, if we follow the techniques used in \cite[Theorem 4.2]{huang2012social}, 
then by the convexity of the running cost, (\ref{decomp}) implies
\iftoggle{jou}{
\begin{align} \label{rep}
&\frac{1}{N}J_{soc}\left(u,x^{(N)}\right) \geq
\frac{1}{N}J_{soc}\left(\hat{u}^{(N)},\hat{x}^{(N)}\right)+
\frac{1}{N}\sum_{i=1}^N \int_0^T r_i
\tilde{u}_i^T 
\hat{u}_i \mathrm{dt} \nonumber\\
&+ \frac{1}{N}\sum_{i=1}^N \int_0^T q
\left( \tilde{x}_i \right)^T \left( 
\hat{x}_i + L \bar{x}\right)\mathrm{dt}
+\epsilon_N \nonumber\\
&+ \frac{1}{N}\sum_{i=1}^N 
\min\limits_{j=1,\dots,l}
\frac{M_{ij}}{2}\|x_i(T)-p_j\|^2 \\
&- \frac{1}{N}\sum_{i=1}^N  
\min\limits_{j=1,\dots,l}\frac{M_{ij}}{2}
\|\hat{x}_i(T)-p_j\|^2 .\nonumber
\end{align}}{
\begin{align} \label{rep}
&\frac{1}{N}J_{soc}\left(u,x^{(N)}\right) \geq
\frac{1}{N}J_{soc}\left(\hat{u}^{(N)},\hat{x}^{(N)}\right)+
\frac{1}{N}\sum_{i=1}^N \int_0^T
\left \{
r_i
\tilde{u}_i^T 
\hat{u}_i 
+  q
\left( \tilde{x}_i \right)^T \left( 
\hat{x}_i + L \bar{x}\right) \right \}\mathrm{dt}
+\epsilon_N \nonumber\\
&+ \frac{1}{N}\sum_{i=1}^N 
\min\limits_{j=1,\dots,l}
\frac{M_{ij}}{2}\|x_i(T)-p_j\|^2- \frac{1}{N}\sum_{i=1}^N  
\min\limits_{j=1,\dots,l}\frac{M_{ij}}{2}
\|\hat{x}_i(T)-p_j\|^2 .
\end{align}
}
We have
\begin{align*}
\frac{d}{dt}\tilde{x}_i^T(\Gamma^{\theta_i}_k \hat{x}_i+\beta_k^{\theta_i})=
-r_i
\tilde{u}_i^T 
\hat{u}_i 
- q
\left( \tilde{x}_i \right)^T \left( 
\hat{x}_i + L \bar{x}\right).
\end{align*}
Hence,
\begin{align} \label{rep1}
&\frac{1}{N}J_{soc}\left(u,x^{(N)}\right) \geq
\frac{1}{N}J_{soc}\left(\hat{u}^{(N)},\hat{x}^{(N)}\right)\\
&
+ \frac{1}{N}\sum_{i=1}^N 
\left(\phi_i\left(x_i(T)\right) 
-   
\phi_i\left(\hat{x}_i(T)\right) -
\tilde{x}_i^T(T) \frac{d}{dx}\phi_i\left(\hat{x}_i(T)\right)\right)
+\epsilon_N, \nonumber
\end{align}
where $\phi_i$ is the final cost of agent $i$. 
If the final costs are convex (which
is not the case), then 
(\ref{rep1}) 
implies (\ref{conv_th}). To deal with 
the non-convexity of the final costs, steps (\ref{rep}) and (\ref{rep1}) are replaced by
(\ref{decomp1}), (\ref{decomp2}) and 
Assumption \ref{assum-neg}.
\end{rem}
\subsection{Asymptotic Optimal Social Cost}

In this section, we give an explicit form of the asymptotic per agent optimal social cost 
$\lim\limits_{N \rightarrow \infty} \inf\limits_{u \in U^N} \frac{1}{N}J_{soc}(u,x^{(N)})$.
In the following lemmas, we start by approximating this asymptotic per agent social cost.
 
\begin{lem} \label{1-app}
Under Assumptions \ref{assum-neg}, \ref{assum-P0} and 
\ref{assumption: finite mean 2},
\[
\lim\limits_{N \rightarrow \infty}
\left|\inf\limits_{u \in U^N} 
\frac{1}{N}J_{soc}\left(u,x^{(N)}\right)-
\frac{1}{N}J_{soc}\left(\hat{u}^{(N)},\bar{x}\right)
\right|=0
\]
\end{lem}
\begin{IEEEproof}
We have
\iftoggle{jou}{
\begin{align*}
&\frac{1}{N}J_{soc}\left(\hat{u}^{(N)},\hat{x}
^{(N)}\right)-\frac{1}{N}J_{soc}
\left(\hat{u}^{(N)},\bar{x}\right) \\
=&\int_0^T \frac{q}{2N} \sum_{i=1}^N
\left(
\left\|\hat{x}_i-Z \hat{x}^{(N)} \right \|^2-
\|\hat{x}_i-Z \bar{x} \|^2
\right )\mathrm{dt}\\
=&\frac{q}{2}\int_0^T \left\|Z 
\left(\hat{x}^{(N)}-\bar{x}\right)\right\|
^2\mathrm{dt}\\
&+q\int_0^T 
\left(\hat{x}^{(N)}-Z \bar{x}\right)^T
Z \left(\bar{x}-\hat{x}^{(N)}\right)
\mathrm{dt}.
\end{align*}}{
\begin{align*}
&\frac{1}{N}J_{soc}\left(\hat{u}^{(N)},\hat{x}
^{(N)}\right)-\frac{1}{N}J_{soc}
\left(\hat{u}^{(N)},\bar{x}\right) 
=\int_0^T \frac{q}{2N} \sum_{i=1}^N
\left(
\left\|\hat{x}_i-Z \hat{x}^{(N)} \right \|^2-
\|\hat{x}_i-Z \bar{x} \|^2
\right )\mathrm{dt}\\
=&\frac{q}{2}\int_0^T \left\|Z 
\left(\hat{x}^{(N)}-\bar{x}\right)\right\|
^2\mathrm{dt}+q\int_0^T 
\left(\hat{x}^{(N)}-Z \bar{x}\right)^T
Z \left(\bar{x}-\hat{x}^{(N)}\right)
\mathrm{dt}.
\end{align*}
}
The Cauchy-Schwarz inequality and Lemma \ref{appr} imply
\[
\lim\limits_{N\rightarrow\infty}
\left|\frac{1}{N}J_{soc}\left(\hat{u}^{(N)},\hat{x}
^{(N)}\right)-\frac{1}{N}J_{soc}
\left(\hat{u}^{(N)},\bar{x}\right)\right|=0.
\] 
Therefore, we deduce by Theorem \ref{th-optim} the result.
\end{IEEEproof} 
\begin{lem} \label{2-app}
Under Assumptions \ref{assum-neg}, \ref{assum-P0} and
\ref{assumption: finite mean 2},
\[
\lim\limits_{N \rightarrow \infty}
\left|\inf\limits_{u \in U^N} 
\frac{1}{N}J_{soc}\left(u,x^{(N)}\right)-
J_{soc}^{\infty}(\bar{x})
\right|=0,
\]
where
\begin{multline*}
J_{soc}^{\infty}(\bar{x})
=\int \Bigg [ \int_0^T 
 \bigg \{ \frac{q}{2} 
 \left\|\hat{x}(t,x^0,\theta)-Z 
 \bar{x}\right\|^2 +\\ 
 \frac{r_{\theta}}{2} \|
 \hat{u}(t,x^0,\theta) \|^2 
 \bigg \} 
 \mathrm{dt} 
 +\min\limits_{j=1,\dots,l}
 \frac{M_{\theta j}}{2} \|
 \hat{x}(T,x^0,\theta)-p_j\|^2
 \Bigg ]
 \mathrm{d} \mathbb{P}_0
 \mathrm{d} \mathbb{P}_\theta.  
\end{multline*}
\end{lem}
\begin{IEEEproof}
By Lemma \ref{1-app}, it suffices to prove that
\[
\lim\limits_{N \rightarrow \infty}
\left|J_{soc}^{\infty}(\bar{x})-
\frac{1}{N}J_{soc}
\left(\hat{u}^{(N)},\bar{x}\right)
\right|=0.
\]
We use the same notation as in the proof of Lemma \ref{appr}. We have
\[
J_{soc}^{\infty}(\bar{x})-
\frac{1}{N}J_{soc}
\left(\hat{u}^{(N)},\bar{x}\right)=\psi_1+\psi_2+\psi_3\]
where
\iftoggle{jou}{
\begin{align*}
\psi_1=&\frac{q}{2}\int_0^T  \int   
 \bigg \{ \left\|\hat{x}\left(t,X^0,\xi^\theta\right)-Z 
 \bar{x} \right\|^2 \\
 &-  \left\|
 \hat{x}\left(t,X^0_N,\xi^\theta_N\right)-Z 
 \bar{x}\right\|^2 \bigg \}
\mathrm{d} \mathbb{P}
 \mathrm{dt}\\
 \psi_2=&\int_0^T  \int \bigg \{  
 \frac{r_{\xi^\theta}}{2} 
 \left
 \|\hat{u}\left(t,X^0,\xi^\theta\right)
 \right\|^2\\
 &-\frac{r_{\xi^\theta_N}}{2} 
\left \|\hat{u}\left(t,X^0_N,\xi^\theta_N\right)
\right\|^2 
 \bigg \}
 \mathrm{d} 
 \mathbb{P} 
 \mathrm{dt}\\
\psi_3=&\int \min\limits_{j=1,\dots,l}
 \frac{M_{\xi^\theta j}}{2} \left\|
 \hat{x}\left(T,X^0,\xi^\theta\right)-p_j
 \right\|^2 
 \mathrm{d} 
 \mathbb{P} \\
 -&\int \min\limits_{j=1,\dots,l}
 \frac{M_{\xi^\theta_N j}}{2} \left \|
 \hat{x}\left(T,X^0_N,\xi^\theta_N\right)-p_j
 \right\|^2
 \mathrm{d} 
 \mathbb{P}.
 \end{align*}}{
 \begin{align*}
\psi_1=&\frac{q}{2}\int_0^T  \int   
 \bigg \{ \left\|\hat{x}\left(t,X^0,\xi^\theta\right)-Z 
 \bar{x} \right\|^2 -  \left\|
 \hat{x}\left(t,X^0_N,\xi^\theta_N\right)-Z 
 \bar{x}\right\|^2 \bigg \}
\mathrm{d} \mathbb{P}
 \mathrm{dt}\\
 \psi_2=&\int_0^T  \int \bigg \{  
 \frac{r_{\xi^\theta}}{2} 
 \left
 \|\hat{u}\left(t,X^0,\xi^\theta\right)
 \right\|^2-\frac{r_{\xi^\theta_N}}{2} 
\left \|\hat{u}\left(t,X^0_N,\xi^\theta_N\right)
\right\|^2 
 \bigg \}
 \mathrm{d} 
 \mathbb{P} 
 \mathrm{dt}\\
\psi_3=&\int \min\limits_{j=1,\dots,l}
 \frac{M_{\xi^\theta j}}{2} \left\|
 \hat{x}\left(T,X^0,\xi^\theta\right)-p_j
 \right\|^2 
 \mathrm{d} 
 \mathbb{P} 
 -\int \min\limits_{j=1,\dots,l}
 \frac{M_{\xi^\theta_N j}}{2} \left \|
 \hat{x}\left(T,X^0_N,\xi^\theta_N\right)-p_j
 \right\|^2
 \mathrm{d} 
 \mathbb{P}.
 \end{align*}
 }
Noting that $a^Ta-b^Tb=(a+b)^T(a-b)$ and that the minimum of $l$ continuous 
functions is continuous, one can prove by the same techniques used in the 
proof of Lemma \ref{appr} that $\psi_1$, $\psi_2$ and $\psi_3$ converge to
zero as $N$ goes to infinity.
\end{IEEEproof}
In the following theorem, we give an explicit form of
the asymptotic social cost. 
This expression depends only on the distributions $\mathbb{P}_0$,
$\mathbb{P}_\theta$ and a fixed point path $\bar{x}$.
\begin{thm} \label{asym-cost-1}
Under Assumptions \ref{assum-neg},
\ref{assum-P0} and 
\ref{assumption: finite mean 2},
\begin{multline*}
\lim\limits_{N \rightarrow \infty}
\inf\limits_{u \in U^N} 
\frac{1}{N}J_{soc}\left(u,x^{(N)}\right)=
 -\frac{1}{2}\int_0^T q
\bar{x}^T L\bar{x} \mathrm{dt}+\\
\sum_{j=1}^l 
\int  \mathbbm{1}_{D^{\theta}_{j}
(\bar{x})}
(x^0)
\Big\{ \frac{1}{2} 
(x^0)^T \Gamma_j^\theta (0) x^0
+(\beta_j^\theta (0))^T x^0 + 
\delta_j^\theta (0) \Big \} \mathrm{d} \mathbb{P}_0 
\mathrm{d} \mathbb{P}_\theta.
\end{multline*}
\end{thm}
\begin{IEEEproof}
Following Lemma \ref{2-app}, the per agent asymptotic optimal social cost
is equal to $J_{soc}^{\infty} (\bar{x})$. Noting (\ref{additional-MFE}), one 
can write $J_{soc}^{\infty}(\bar{x})=\psi_4-\frac{1}{2}\int_0^T q\bar{x}^T L\bar{x} \mathrm{dt}$,
where
\iftoggle{jou}{
\begin{multline*}
\psi_4=
\int \Bigg [ \int_0^T \bigg\{ 
\frac{q}{2}
\left\|\hat{x}
\left(t,x^0,\theta\right)\right\|^2 + q 
\bar{x}^{T} L \hat{x}
\left(t,x^0,\theta\right) \\
+ \frac{r_\theta}{2} 
\left\|\hat{x}
\left(t,x^0,\theta\right) \right\|^2 \bigg\} 
\mathrm{dt} \\
 +\min\limits_{j=1,\dots,l}
 \frac{M_{\theta j}}{2} \left\|
 \hat{x}
\left(T,x^0,\theta\right)-p_j \right\|^2 \Bigg ]
 \mathrm{d} \mathbb{P}_0 
 \mathrm{d} \mathbb{P}_\theta=\\
 \sum_{j=1}^l 
\int  \mathbbm{1}_{D^{\theta}_{j}
(\bar{x})}
(x^0)
\Big\{ 
(x^0)^T \Gamma_j^\theta (0) x^0
+\beta_j^\theta (0)^T x^0 + 
\delta_j^\theta (0) \Big \} \mathrm{d} \mathbb{P}_0 
\mathrm{d} \mathbb{P}_\theta.
\end{multline*}}{
\begin{multline*}
\psi_4=
\int \Bigg [ \int_0^T \bigg\{ 
\frac{q}{2}
\left\|\hat{x}
\left(t,x^0,\theta\right)\right\|^2 + q 
\bar{x}^{T} L \hat{x}
\left(t,x^0,\theta\right) 
+ \frac{r_\theta}{2} 
\left\|\hat{x}
\left(t,x^0,\theta\right) \right\|^2 \bigg\} 
\mathrm{dt} \\
 +\min\limits_{j=1,\dots,l}
 \frac{M_{\theta j}}{2} \left\|
 \hat{x}
\left(T,x^0,\theta\right)-p_j \right\|^2 \Bigg ]
 \mathrm{d} \mathbb{P}_0 
 \mathrm{d} \mathbb{P}_\theta=\\
 \sum_{j=1}^l 
\int  \mathbbm{1}_{D^{\theta}_{j}
(\bar{x})}
(x^0)
\Big\{ 
(x^0)^T \Gamma_j^\theta (0) x^0
+\beta_j^\theta (0)^T x^0 + 
\delta_j^\theta (0) \Big \} \mathrm{d} \mathbb{P}_0 
\mathrm{d} \mathbb{P}_\theta.
\end{multline*}
}
\end{IEEEproof}
\section{Simulation Results} \label{sim}

In this section, we compare numerically the cooperative and the 
non-cooperative behaviors of a group of agents choosing between two alternatives 
under the social effect. We consider a uniform group of $400$ players initially 
drawn from the Gaussian distribution 
$\mathcal{N}\Big(\begin{bmatrix}-5 & 10\end{bmatrix}^T,15 I_2\Big)$ and 
moving in $\mathbb{R}^2$ according to the dynamics
\begin{align*}
A_i=\begin{bmatrix}
0 && 1\\
0.02 && -0.3
\end{bmatrix} &&
B_i=\begin{bmatrix}
0\\
0.3
\end{bmatrix}
\end{align*}
towards the potential destination points $p_{1}=(-10,0)$ or $p_{2}=(10,0)$.
Hence we have a binary choice problem, and in this case one can characterize the way 
the population splits between the alternatives, in both the cooperative and non-cooperative 
cases, by a number $\lambda$, which is the fraction of players that go towards $p_1$. 
This number $\lambda$ is a fixed point of a well defined function $F$ and can be 
computed by dichotomy. Moreover, one can compute the fixed point path $\bar{x}$ that 
corresponds to $\lambda$. For more details one can refer to
\cite[Theorem 6]{DBLP:journals/corr/SalhabMN15} and 
\cite[Section 5.A]{DBLP:journals/corr/SalhabMN15}. We set $r_i=10$, $M_{ij}=1200$, $T=2$, $Z=3.5I_2$, 
and we vary the social effect coefficient $q$.
$L=Z^TZ-Z-Z^T=5.25I_2$ satisfies Assumption \ref{assum-neg}.
For $q=0$ (no social effect), Fig. \ref{Fig.1} and \ref{Fig.2} show that the 
$82\%$ of the players (green squares in Fig. \ref{Fig.2}) go towards $p_2$ 
in both the cooperative and non-cooperative cases. 
As the social effect increases ($q$ increases from $0$
to $45$), in the non-cooperative case, the 
majority influences the minority whose size reduces from $18\%$ to zero 
(Fig. \ref{Fig.1} and \ref{Fig.4}).
In the cooperative case however, 
the size of the majority decreases and the population splits more
evenly between the two choices (Fig. \ref{Fig.1} and \ref{Fig.3}). 
Fig. \ref{Fig.1} also illustrates that the per agent social cost in the cooperative
case is smaller than in the non-cooperative case.
\iftoggle{jou}{
\begin{figure}[h!]
    \centering
 \includegraphics[width=0.49\textwidth]{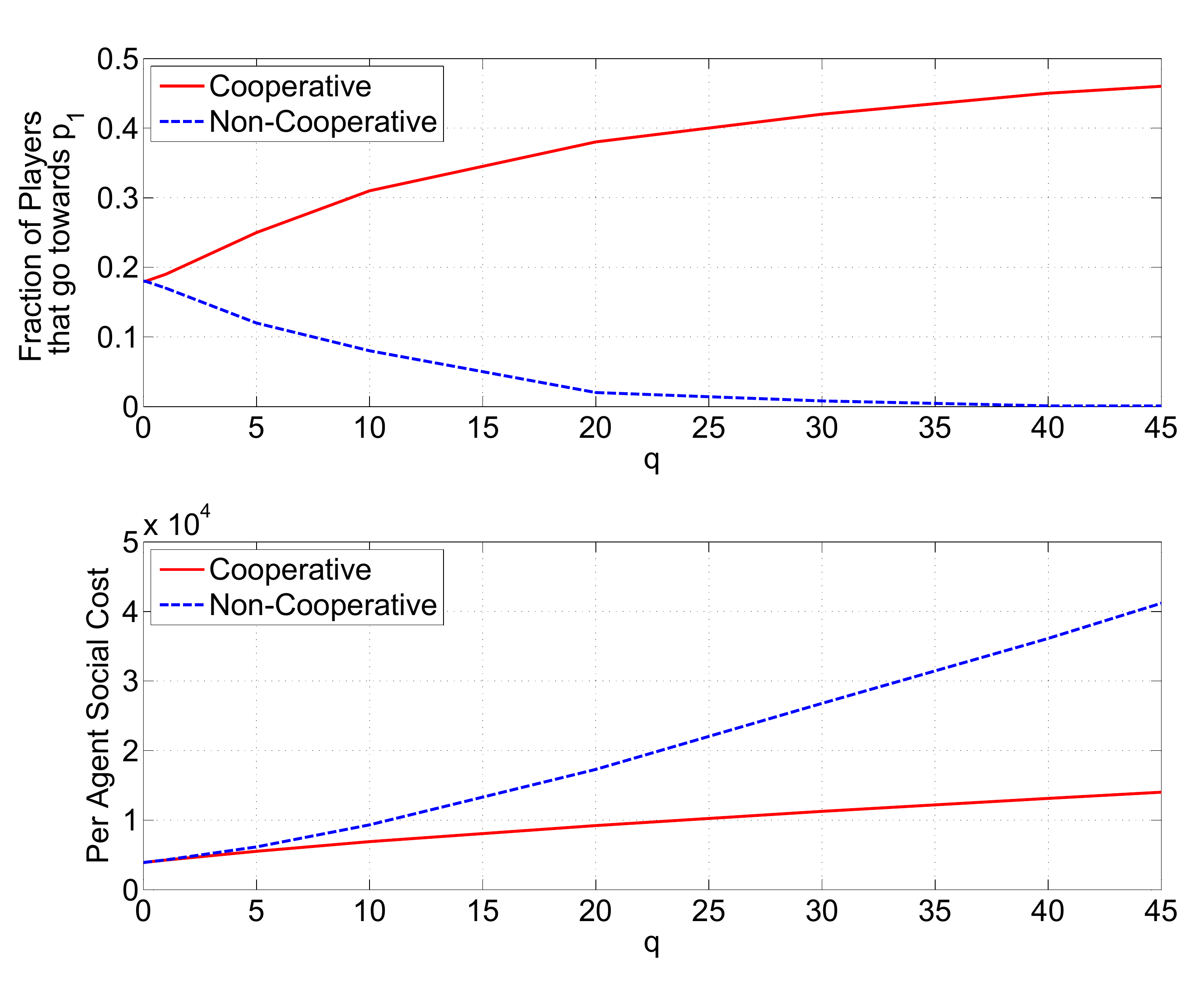}
    \caption{Cooperative vs. non-cooperative behavior}
    \label{Fig.1}
\end{figure}
\begin{figure}[h!]
    \centering
 \includegraphics[width=0.49\textwidth]{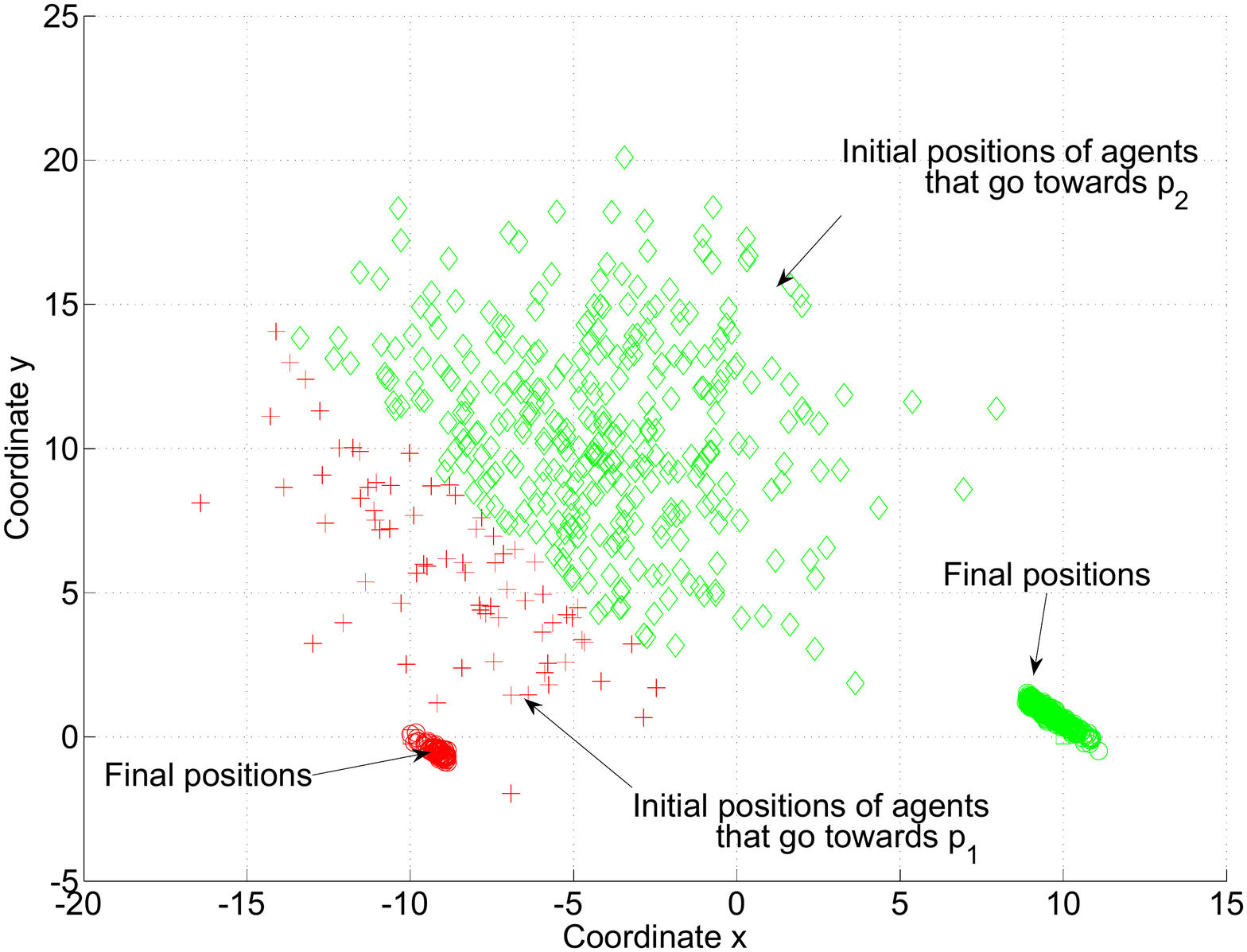}
    \caption{Absence of social effect ($q=0$). The majority goes towards $p_2$.}
    \label{Fig.2}
\end{figure}
\begin{figure}[h!]
    \centering
 \includegraphics[width=0.49\textwidth]{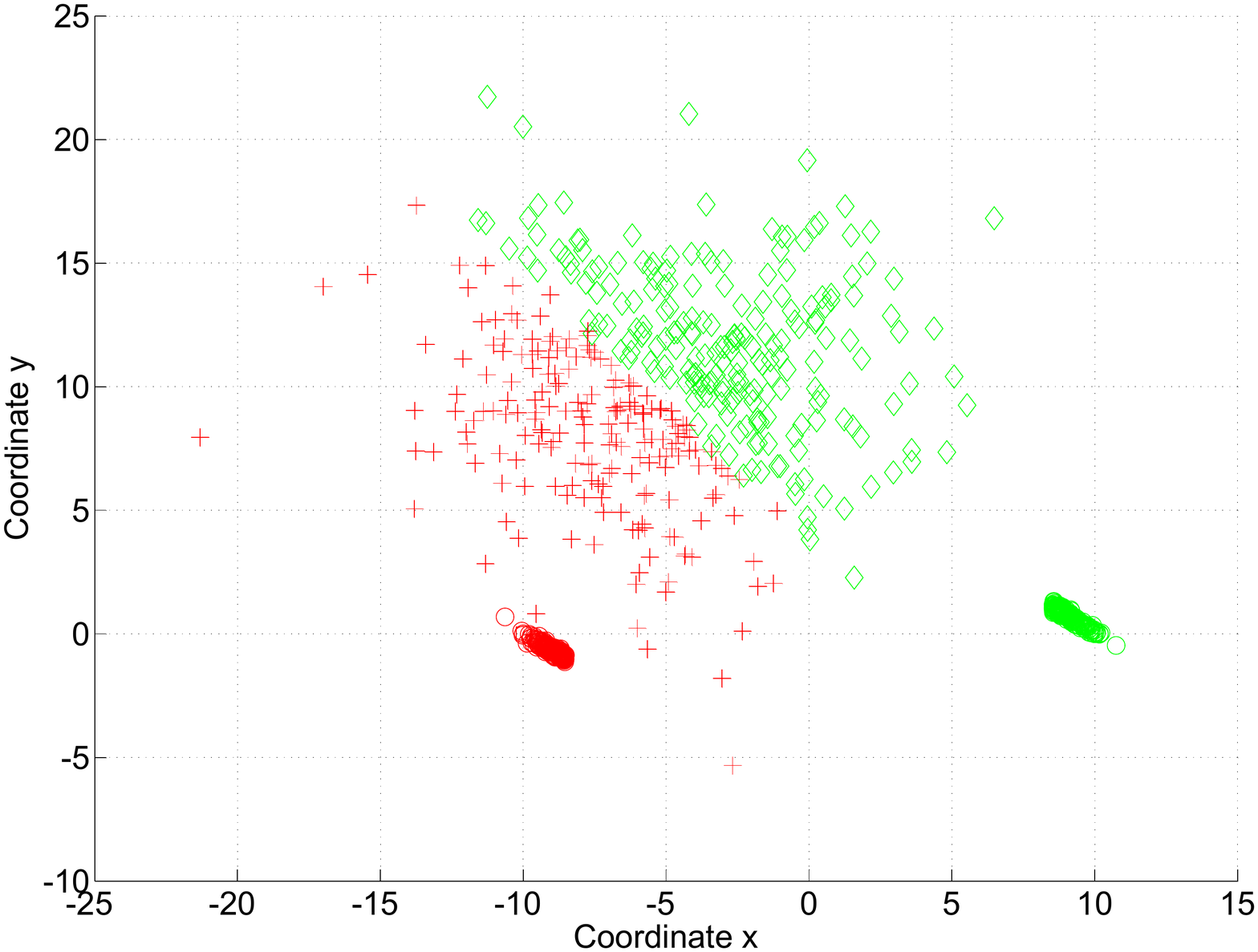}
    \caption{Cooperative case with high social effect ($q=40$). 
    The population splits more evenly.}  
    \label{Fig.3}
\end{figure}
\begin{figure}[h!]
    \centering
 \includegraphics[width=0.49\textwidth]{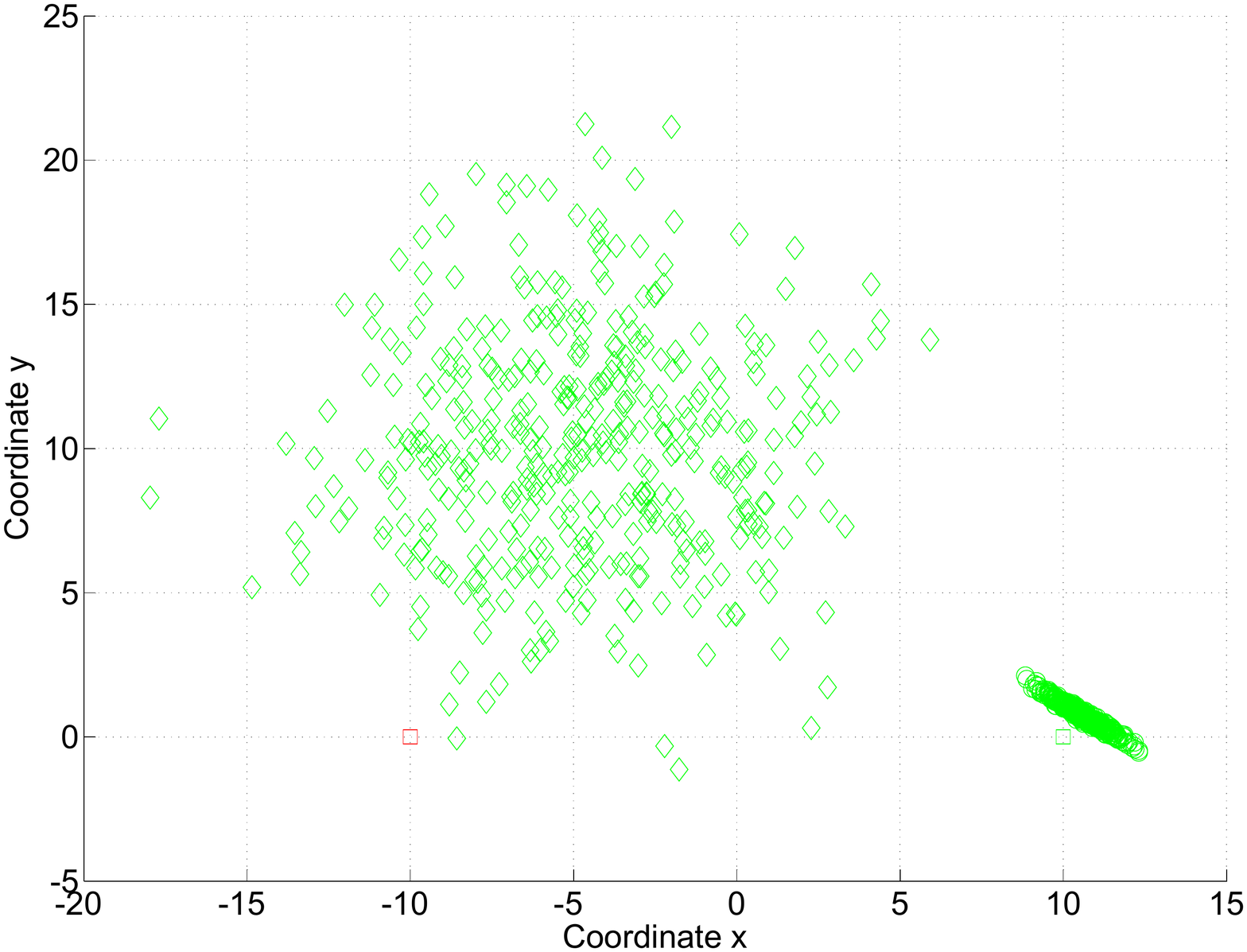}
    \caption{Non-cooperative case with high social effect ($q=40$). 
    The population reaches consensus on $p_2$.
    }
    \label{Fig.4}
\end{figure}}{
\begin{figure}[h!]
    \centering
 \includegraphics[width=0.7\textwidth]{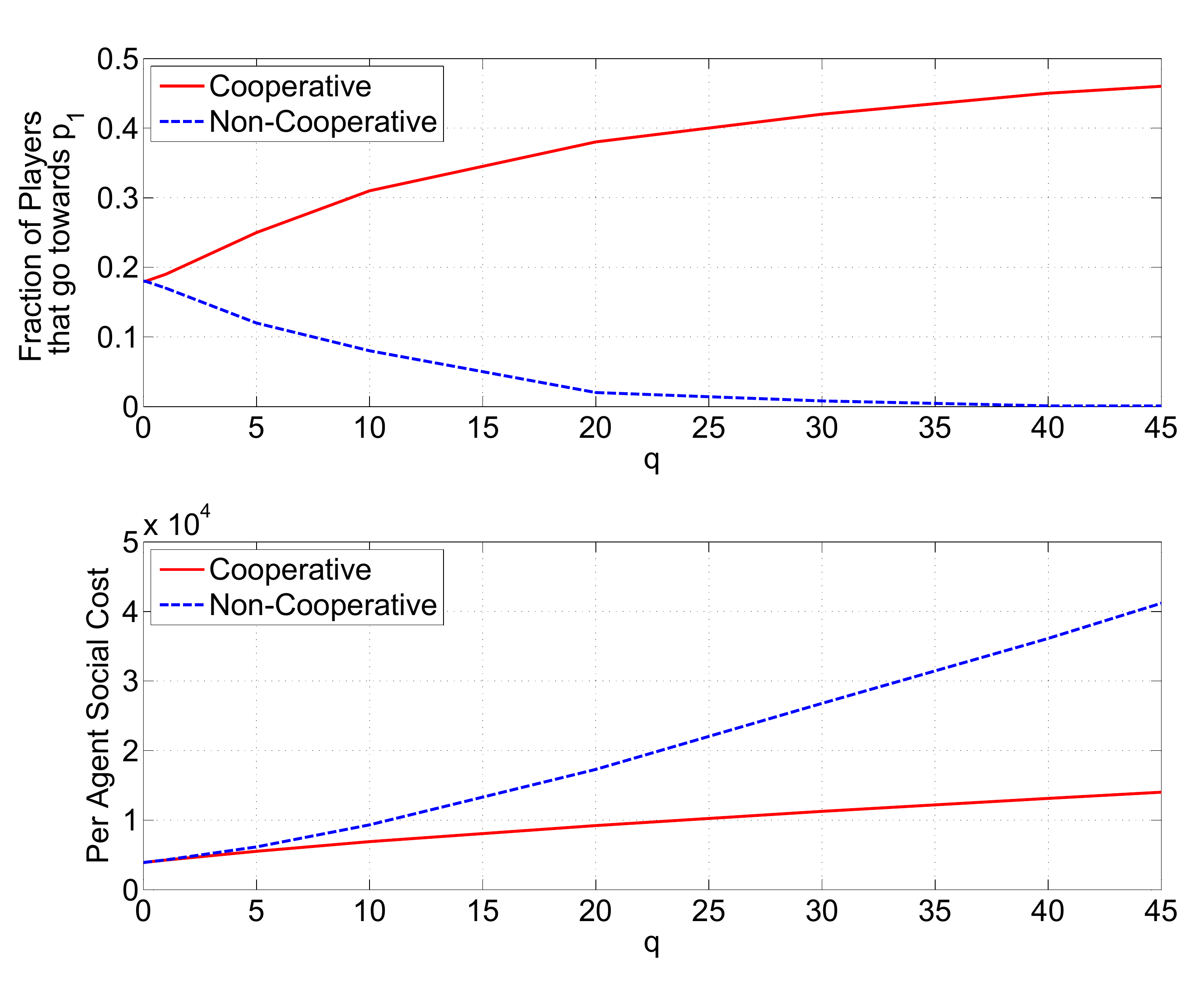}
    \caption{Cooperative vs. non-cooperative behavior}
    \label{Fig.1}
\end{figure}
\begin{figure}[h!]
    \centering
 \includegraphics[width=0.7\textwidth]{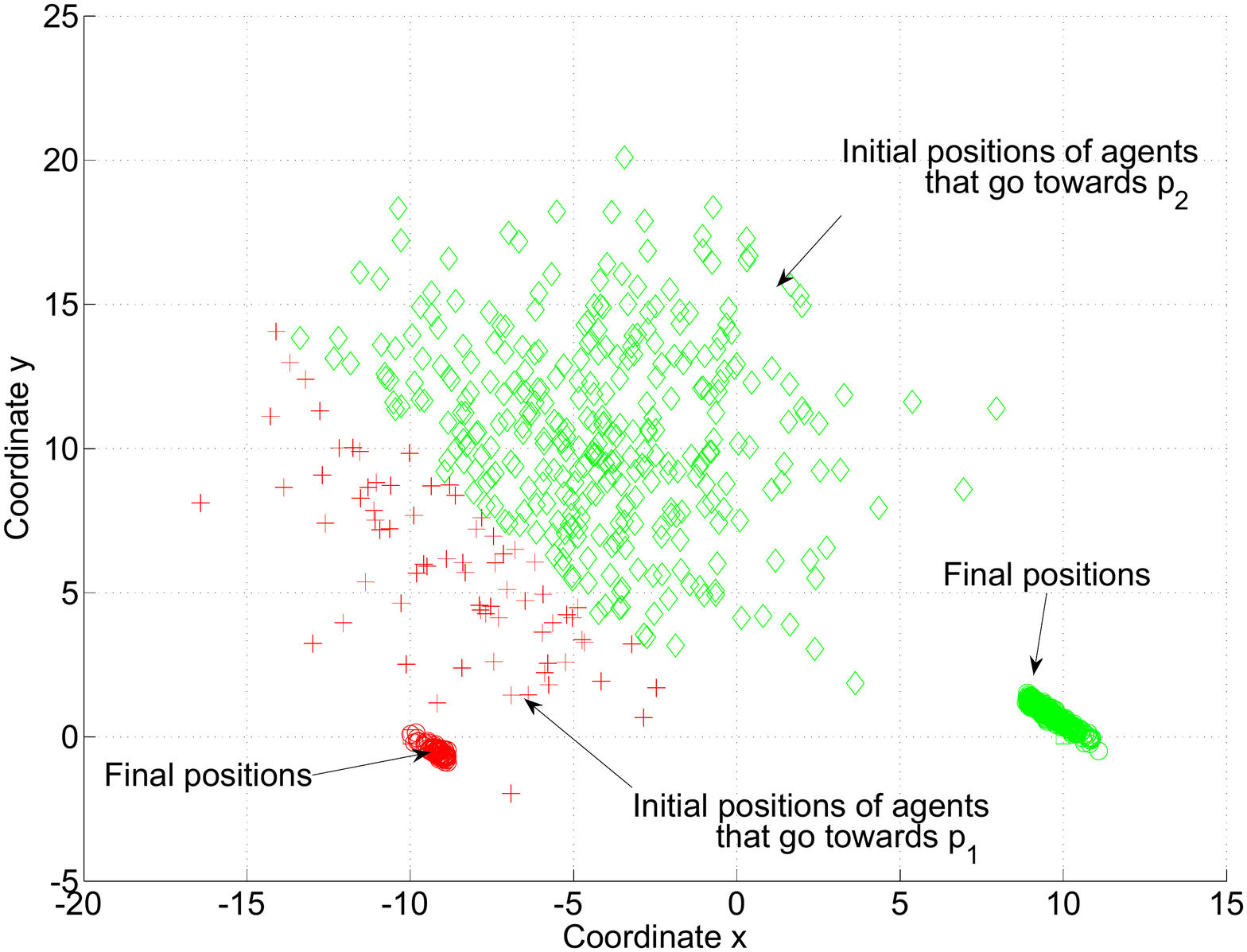}
    \caption{Absence of social effect ($q=0$). The majority goes towards $p_2$.}
    \label{Fig.2}
\end{figure}
\begin{figure}[h!]
    \centering
 \includegraphics[width=0.7\textwidth]{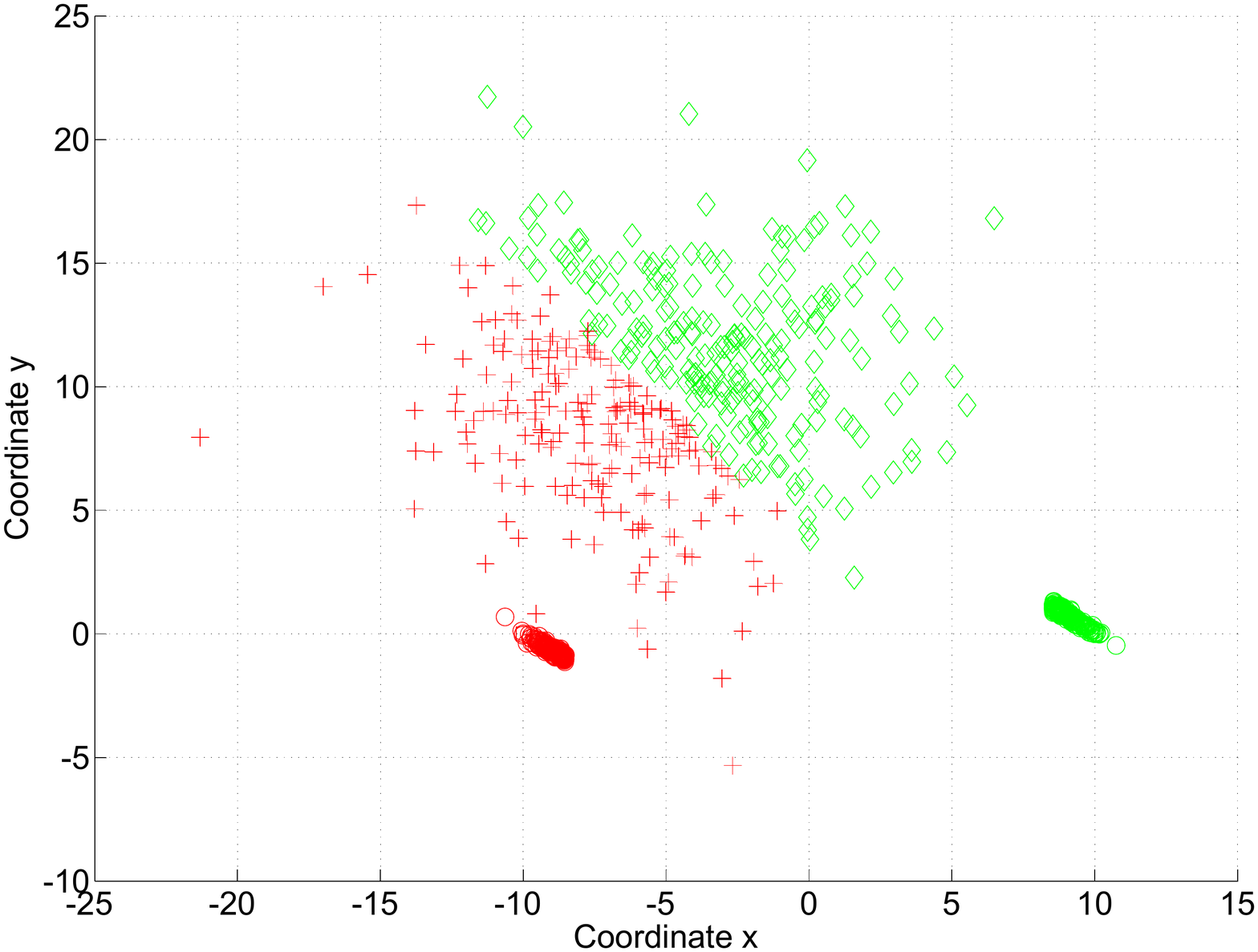}
    \caption{Cooperative case with high social effect ($q=40$). 
    The population splits more evenly.}  
    \label{Fig.3}
\end{figure}
\begin{figure}[h!]
    \centering
 \includegraphics[width=0.7\textwidth]{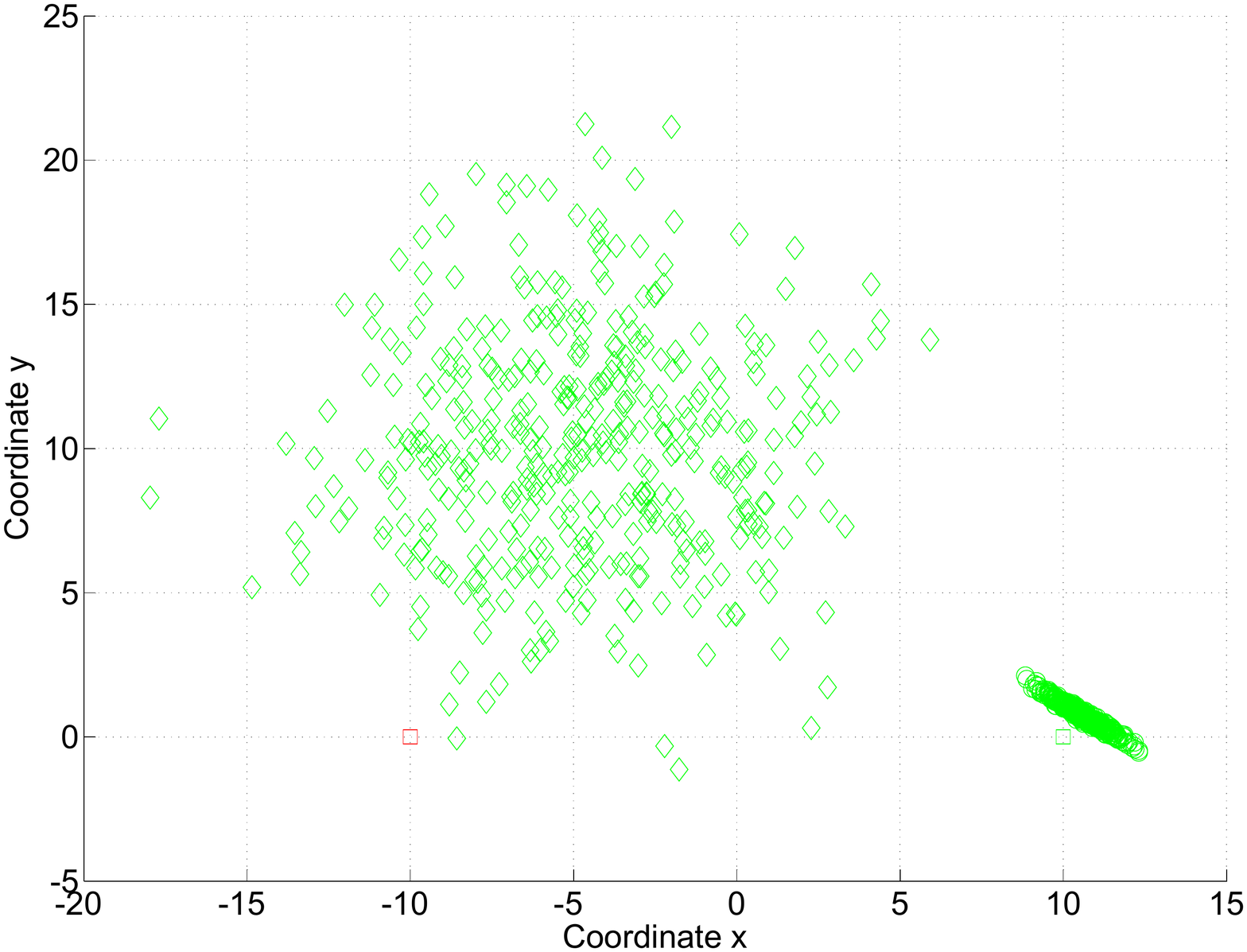}
    \caption{Non-cooperative case with high social effect ($q=40$). 
    The population reaches consensus on $p_2$.
    }
    \label{Fig.4}
\end{figure}
}

\section{Conclusion} \label{conclusion}

We consider in this paper a dynamic cooperative game model where a large 
number of players are making a socially influenced choice between multiple 
alternatives. Finding
an exact social optimum can be done by solving a number of LQR problems 
that grows exponentially with the number of players.
Alternatively, we develop via the MFG methodology a set of decentralized 
strategies that are asymptotically socially optimal. The computation of the 
decentralized strategies assumes that each agent knows the statistical 
distributions of the initial states and parameters. 
For future work, it is of interest to consider situations where the
cooperative players learn these statistical distributions while moving 
towards the destination points, e.g., by sharing and updating their current 
states and parameters through a random communication graph.

\ifCLASSOPTIONcaptionsoff
  \newpage
\fi

\bibliographystyle{IEEEtran}
\bibliography{IEEEabrv,mfg}

\end{document}